\documentstyle[aps,preprint,pra,url]{revtex}
\begin{document}
\newcommand{\etal}{{\em et al.}\/}
\newcommand{\IP}{inner polarization}
\newcommand{\IPF}{\IP\ function}
\newcommand{\IPFs}{\IP\ functions}
\newcommand{\auth}[2]{#1 #2, }
\newcommand{\jcite}[4]{#1 {\bf #2}, #3 (#4)}
\newcommand{\et}{ and }
\newcommand{\twoauth}[4]{#1 #2 and #3 #4,}
\newcommand{\oneauth}[2]{#1 #2,}
\newcommand{\andauth}[2]{and #1 #2, }
\newcommand{\book}[4]{{\it #1} (#2, #3, #4)}
\newcommand{\inbook}[5]{In {\it #1} (ed. #2), #3, #4, #5}
\newcommand{\erratum}[3]{\jcite{erratum}{#1}{#2}{#3}}
\newcommand{\JCP}[3]{\jcite{J. Chem. Phys.}{#1}{#2}{#3}}
\newcommand{\jms}[3]{\jcite{J. Mol. Spectrosc.}{#1}{#2}{#3}}
\newcommand{\jmsp}[3]{\jcite{J. Mol. Spectrosc.}{#1}{#2}{#3}}
\newcommand{\jmstr}[3]{\jcite{J. Mol. Struct.}{#1}{#2}{#3}}
\newcommand{\theochem}[3]{\jcite{J. Mol. Struct. (Theochem)}{#1}{#2}{#3}}
\newcommand{\cpl}[3]{\jcite{Chem. Phys. Lett.}{#1}{#2}{#3}}
\newcommand{\cp}[3]{\jcite{Chem. Phys.}{#1}{#2}{#3}}
\newcommand{\pr}[3]{\jcite{Phys. Rev.}{#1}{#2}{#3}}
\newcommand{\CR}[3]{\jcite{Chem. Rev.}{#1}{#2}{#3}}
\newcommand{\jpc}[3]{\jcite{J. Phys. Chem.}{#1}{#2}{#3}}
\newcommand{\jpcA}[3]{\jcite{J. Phys. Chem. A}{#1}{#2}{#3}}
\newcommand{\jpca}[3]{\jcite{J. Phys. Chem. A}{#1}{#2}{#3}}
\newcommand{\jpcB}[3]{\jcite{J. Phys. Chem. B}{#1}{#2}{#3}}
\newcommand{\PRA}[3]{\jcite{Phys. Rev. A}{#1}{#2}{#3}}
\newcommand{\PRB}[3]{\jcite{Phys. Rev. B}{#1}{#2}{#3}}
\newcommand{\jcc}[3]{\jcite{J. Comput. Chem.}{#1}{#2}{#3}}
\newcommand{\molphys}[3]{\jcite{Mol. Phys.}{#1}{#2}{#3}}
\newcommand{\mph}[3]{\jcite{Mol. Phys.}{#1}{#2}{#3}}
\newcommand{\cpc}[3]{\jcite{Comput. Phys. Commun.}{#1}{#2}{#3}}
\newcommand{\jcsfii}[3]{\jcite{J. Chem. Soc. Faraday Trans. II}{#1}{#2}{#3}}
\newcommand{\prsa}[3]{\jcite{Proc. Royal Soc. A}{#1}{#2}{#3}}
\newcommand{\jacs}[3]{\jcite{J. Am. Chem. Soc.}{#1}{#2}{#3}}
\newcommand{\ijqcs}[3]{\jcite{Int. J. Quantum Chem. Symp.}{#1}{#2}{#3}}
\newcommand{\ijqc}[3]{\jcite{Int. J. Quantum Chem.}{#1}{#2}{#3}}
\newcommand{\spa}[3]{\jcite{Spectrochim. Acta A}{#1}{#2}{#3}}
\newcommand{\tca}[3]{\jcite{Theor. Chem. Acc.}{#1}{#2}{#3}}
\newcommand{\tcaold}[3]{\jcite{Theor. Chim. Acta}{#1}{#2}{#3}}
\newcommand{\jpcrd}[3]{\jcite{J. Phys. Chem. Ref. Data}{#1}{#2}{#3}}
\newcommand{\arpc}[3]{\jcite{Ann. Rev. Phys. Chem.}{#1}{#2}{#3}}

\draft
\title{Correlation consistent valence basis sets for use with
the Stuttgart-Dresden-Bonn relativistic effective core potentials:
the atoms Ga--Kr and In-Xe.}
\author{Jan M.L. Martin* and Andreas Sundermann}
\address{Department of Organic Chemistry,
Kimmelman Building, Room 262,
Weizmann Institute of Science,
IL-76100 Re\d{h}ovot, Israel. {\rm E-mail:} {\tt comartin@wicc.weizmann.ac.il}
}
\date{{\it J. Chem. Phys.} MS A0.09.107; Received Sept. 14, 2000; Revised 
\today}
\maketitle
\begin{abstract}
We propose large-core correlation-consistent pseudopotential basis sets for
the heavy p-block elements Ga--Kr and In--Xe. The basis sets are 
of cc-pVTZ and cc-pVQZ quality, and have been optimized for use with
the large-core (valence-electrons only)
Stuttgart-Dresden-Bonn relativistic pseudopotentials.
Validation calculations on a variety of third-row and fourth-row 
diatomics suggest them to be comparable in quality to the all-electron
cc-pVTZ and cc-pVQZ basis sets for lighter elements. Especially the
SDB-cc-pVQZ basis set in conjunction with a core polarization potential (CPP)
yields excellent agreement with experiment for compounds of the later 
heavy p-block elements. For accurate calculations on Ga (and, to a 
lesser extent, Ge) compounds,
explicit treatment of 13 valence electrons appears to be desirable,
while it seems inevitable for In compounds. For Ga and Ge, we propose
correlation consistent basis sets extended for (3d) correlation.
For 
accurate calculations on
organometallic complexes of interest to homogenous catalysis, we
recommend a combination of the standard cc-pVTZ basis set for first- and
second-row elements, the presently derived SDB-cc-pVTZ basis set for
heavier p-block elements, and for transition metals,
the small-core [6s5p3d] Stuttgart-Dresden
basis set-RECP combination supplemented by $(2f1g)$ functions with exponents
given in the Appendix to the present paper.
\end{abstract}


\section{Introduction and theoretical background}

The two major factors that determine the quality of a wavefunction-based 
electronic structure calculation are the quality of the one-particle
basis set and that of the n-particle correlation treatment.

Thanks to great progress in electron correlation methods
(notably in the area of coupled cluster theory\cite{ccreviews}), the n-particle
problem is to a large extent solved, leaving the 1-particle basis set
as the main factor that determines the quality of an electronic structure
calculation.

Abundant research has been carried out on basis set convergence and the
development of extended basis sets for first- and second-row systems
(see e.g.\cite{Hel95} for a review): we note in particular the ANO
(atomic natural orbital\cite{Alm87}) basis sets of Alml\"of and Taylor,
the WMR (Widmark-Malmqvist-Roos, or averaged ANO\cite{WMR}) basis sets
of the eponymous group,
and the correlation consistent ($cc$) basis sets of Dunning and coworkers\cite{Dun89,DunECC}.
Due to their relative compactness in terms of Gaussian primitives, the $cc$
basis sets have become very popular for benchmark wavefunction-based ab initio
calculations: to a lesser extent, the same holds true for DFT (density functional
theory\cite{Par89}) calculations.

Basis set convergence of the dynamical correlation energy in conventional 
electronic structure calculations is known to be very slow. This is less of
an issue for DFT calculations\cite{And97,MarVUB,DePVUB,ip,quad}: as a rule
basis set convergence appears to be reached for basis sets of $spdf$ 
quality and certainly for basis sets of $spdfg$ quality. Standard
basis sets of such quality are readily available for first- and second-row
compounds: in addition, ANO and WMR basis sets are available for the
first-row transition metals\cite{Bau95} and $cc$ basis sets for the third-row main
group elements\cite{Wil99}.

Our group has recently become involved in a number of mechanistic studies
by means of DFT methods (e.g. on competitive CC/CH activation by Rh(I) 
pincer complexes\cite{Rh-PCP,Rh-PCN} and on Pd(0/II) and Pd(II/IV)
catalyzed mechanisms of the Heck reaction\cite{Heck}) that involve
second-row transition metals and fourth-row main group elements.
Generally, one is limited to basis set/ECP (effective core potential)
combinations of approximately valence double-zeta quality. If one
wants to establish basis set convergence for a given property, one
is forced to optimize basis sets ad hoc (as we have done\cite{Heck}), 
which is however not necessarily the most elegant solution.
Given that present-day DFT methods are less than ideal for the treatment
of transition states\cite{Bak95,Tru2K,sn2}, calibration calculations
using coupled cluster methods are in order (at least for some small
model systems) --- and here the basis set issue becomes even more important.

It is well known that
for such heavy elements, relativistic effects cannot gratuitously
be neglected without paying a heavy toll in terms of reliability.
The theory of relativistic electronic structure methods has been
reviewed in detail by Pyykk\"o \cite{Pyk88} and most recently
by Reiher and Hess\cite{Hess2000}. For systems in the size range
of interest to organometallic chemists, four-component all-electron
relativistic calculations are presently out of the question, and 
even quasirelativistic calculations are very costly: consequently,
by far the most commonly employed alternative has been the
application of relativistic effective core potentials (RECPs).
A useful `fringe benefit' of the latter is that they reduce the
number of electrons that need to be treated, and hence, indirectly,
the overall size of the basis set and cost of the calculation.

The theory and practice of ECPs have been reviewed repeatedly
(e.g.\cite{KraSteARPC84,Ermler1988,Gropen1988}), most recently
by Dolg\cite{Dolg2000}. Several ECP families are available for the
range of the periodic table of interest to us, such as the
LANL (Los Alamos National Laboratory) ECPs of Hay and Wadt\cite{lanl},
the CEP (Consistent Effective Potential) family of Stevens, Basch, and
coworkers\cite{sbk}, the Ermler-Christensen family\cite{ermler-christensen},
and the Stuttgart-Dresden-Bonn (SDB) energy-consistent pseudopotentials\cite{sddmain}.

The purpose of this paper is to present and validate valence basis sets
for RECPs of a quality comparable to that of the cc-pVTZ and cc-pVQZ
correlation-consistent basis sets for lighter elements, to be used
in conjunction with the latter. In selecting the underlying RECP,
we have opted for the SDB pseudopotentials for the following
methodological and pragmatic reasons (some, but not all, of which
are satisfied for the other popular ECPs):
\begin{itemize}
\item compact mathematical form
\item ready availability in the commonly used quantum chemistry
packages Gaussian 98 \cite{g98} and MOLPRO 2000\cite{m2k}
\item consistent treatment of relativistic effects in all relevant
rows of the periodic table
\item independence of the ECP on the valence basis set
\item availability of core polarization potentials (CPPs)\cite{cpp}, since we were
planning to use `large core' potentials for the main group elements
\item availability of extended valence basis sets (specifically, [6s5p3d]
contractions) for the transition
metals. In Appendix I, we shall present optimized $[2f1g]$ polarization
functions for those valence basis sets, to be used in conjunction with
the presently derived SDB-cc-pVTZ basis sets for third- and fourth-row
elements, and standard cc-pVTZ basis sets for first-and second-row
elements.
\end{itemize}
To our knowledge, the only published example so far of a `correlation consistent' 
basis set based on an ECP is the work of Bauschlicher\cite{Bau99In}, who
published cc-pV$n$Z ($n=$T,Q,5) basis sets for indium, optimized for $(5s,5p,4d)$ correlation,
to be used in conjunction with a small-core SDB pseudopotential. In this
paper and a subsequent application study\cite{Bau99InBIS}, benchmark
calculations on a number of In compounds were presented that clearly
support the idea that the development of SDB-based correlation consistent
basis sets is warranted.

In the next section, we shall describe the procedure by which the valence
basis sets were optimized. In the following section, we shall present
validation calculations with these basis sets on a variety of diatomic
molecules. Conclusions are presented in a final section.

\section{Generation of basis sets}

All electronic structure calculations were carried using MOLPRO2000\cite{m2k}
running on a Compaq ES40 at the Weizmann Institute of Science. Basis sets were
carried out by means of an adaptation of the DOMIN program by P. Spellucci\cite{domin},
which is an implementation of the BFGS (Broyden-Fletcher-Goldfarb-Shanno) 
variable-metric method. Numerical derivatives of order two, four, and six were
used: the lower orders until an approximate minimum was reached, after which
the optimization was refined using the higher orders.

For the third-row main group elements, we employed the SDB pseudopotentials
denoted by the SDB group as ECP28MWB\cite{Ber93}, i.e. large-core (1s2s2p3s3p3d) 
energy-consistent pseudopotentials obtained from quasirelativistic
Wood-Boring\cite{Woo80} calculations. For the fourth-row main group
elements, we employed the ECP46MWB set\cite{Ber93}, i.e. large-core (1s2s2p3s3p3d4s4p4d).

Unless indicated otherwise, all Hartree-Fock calculations were carried out
using proper symmetry and spin eigenfunctions. 

\subsection{Valence $sp$ basis sets}

In order to obtain an idea as to the size of the required $sp$ set for
the valence orbitals, we carried out the following numerical experiment
for the Se atom: a valence SCF calculation was carried out using the complete
(26s17p) part of the all-electron cc-pV5Z basis set\cite{Wil99} added to the ECP28MWB 
pseudopotential. Then all primitives with coefficients below 10$^{-5}$ were
discarded, leaving us with a (16s13p) primitive set at the expense of only
0.38 microhartree in energy. Raising the `cutoff' to 10$^{-4}$ reduced the
primitive set to (13s11p), and raises the energy by another 3 microhartree. 
Raising the cutoff by another order of magnitude reduces the primitive set
to (12s9p), at the expense of an additional 13 microhartree. Applying 
the same sequence of cutoffs to the (21s16p) primitives in the all-electron
cc-pVQZ basis set leads to (14s11p), (13s9p), and (11s7p), respectively:
from the (20s13p) primitives of the all-electron cc-pVTZ basis set we
obtain in the same manner (12s9p) for a 10$^{-4}$ cutoff, and (10s7p) for
a 10$^{-3}$ cutoff. Similar patterns were observed for other third-row
elements: the bottom line appears to be that 3--4 more $s$ primitives are
required than $p$ primitives.

We subsequently attempted to minimize $((k+4)skp)$ basis sets ($k$=6--10)
directly at the SCF level. However, the Hessian for some of the higher-exponent
$s$ functions is extremely flat, and as a result no reliable optimization
can be carried out. Considering the fact that, for instance, the outer (13s11p)
exponents of the all-electron cc-pV5Z basis set display roughly even-tempered
sequences $\zeta_k=\alpha\beta^{k-1}$ {\em except} for the outermost four
primitives of every symmetry, we adopted the compromise solution of 
optimizing the four outermost primitives of each symmetry without restriction,
but constraining the remainder to follow an even-tempered sequence. This leads
to an optimization problem with twelve parameters in all (eight independent
exponents, plus one $\alpha,\beta$ pair each for $s$ and $p$).

In this manner, we were able to obtain (10s6p) through (14s10p) primitive sets.
For Ga, Ge, and As, multiple minima were invariably found, with a solution
that exhibits a `gap' between the 3rd and 4th (or 4th and 5th) outermost 
primitive being marginally lower in energy than a solution where {\it no} such
gaps were present. (This behavior is particularly noticeable for the
$s$ primitives.) Carrying out 4-parameter optimizations with purely
even-tempered (14s10p) basis set quickly reveals the cause: as $\zeta$ increases,
the coefficients are initially positive, but then decay and change sign as the
higher exponent primitives ensure the proper inner shape of the orbital. 
The energy is rather insensitive to the location --- or even the presence ---
of the primitive near the crossing point, and especially with smaller sets of
primitives, a marginal gain in energy might be obtained from a solution with
an additional primitive in the very high exponent region rather than in
the `crossing' region. Since for application in correlated calculations,
the presence of a gap in the outer part of the exponent sequence is 
clearly undesirable, we have deliberately chosen the most `even-tempered'
solution even where it was not the global minimum.

Similar phenomena were observed for In--I: and likewise, we obtained the most
`even-tempered' primitive valence sets up to (14s10p).

\subsection{Addition of higher angular momentum functions}

Parameters for added higher angular momentum functions were then 
optimized at the CISD level. At first even-tempered sequences
of up to four $(3d)$-type functions were added, followed by
up to three additional $(4f)$-type functions and up to two
additional $(5g)$-type functions. For the third-row main group
elements, these optimizations progressed uneventfully. Not 
surprisingly, the $d$ exponents differ somewhat from those
obtained by Dunning and coworkers for all-electron basis sets:
in the latter, the $d$ functions do double-duty as angular 
correlation functions for the $(4s,4p)$ orbitals and as 
$(3d)$ primitives, while in our case they solely take on the former
role. For the $f$ and $g$ functions, the similarity is greater.
In terms of energetic increments, the familiar `correlation 
consistent' (2d1f) and (3d2f1g) groupings of functions with similar 
energy lowerings emerge.

In the fourth row, the convergence pattern of the $d$ exponents is
somewhat peculiar, in that for instance for Te and I, the energy lowerings
for the 2nd and 3rd $(3d)$ function are similar. 
This is caused by the
rather low-lying $(5d)$ orbital, which also causes a somewhat peculiar
$(2d)$ exponent pattern for Te. We shall return to this point shortly.

\subsection{Definition of the final contracted basis sets}

We carried out an analysis similar to that of Dunning and coworkers,
in that we for instance completely contracted the $p$ orbital in a 
$(14s10p4d3f2g)$ basis set, then optimized even-tempered sequences
of added $p$ primitives. The optimum $s$ and $p$ exponents revealed
similar trends. In terms of contracting our Se basis set for 
correlation, however, they unequivocally suggest that the 2nd and 4th
outermost (s) and (p) primitives be decontracted for a valence triple zeta
basis set, and the 2nd--4th outermost primitives for a valence quadruple zeta
basis set. (From here on, we shall be counting primitives starting
from the `outermost', i.e. smallest and most diffuse, exponent.)
By comparison, in the Dunning all-electron case these
were the 1st and 3rd, and 1st--3rd primitives, respectively. However,
our outermost $(sp)$ primitives are considerably more diffuse 
than theirs, by virtue of the absence of the inner-shell 'gravity well'
in the valence-only optimizations. The exponents of the decontracted
primitives in fact are fairly similar.

This having been established, we determined our favored 'VTZ' and 
'VQZ' contraction patterns for each element by comparing total energies 
between all six and
four possible choices, respectively, among the four outermost primitives.
If we denote decontraction of a primitive by a 1 and the lack thereof
by a 0, and start at the lowest exponent, then the favored (i.e., lowest-energy) quadruple-zeta
contraction pattern is found to be \{0111\} for Se, Br, Kr, Te, I, and Xe,
but \{1101\} for Ga, Ge, In, and Sn. (For As, \{1011\} is marginally
lower in energy than \{0111\}, while for Sb, a \{1110\} pattern 
for the $s$ was combined with a \{1101\} pattern for the $p$ functions.)
For the triple-zeta contractions, the \{0101\} pattern prevails for 
As, Se, Br, Kr, Te, I, and Xe, 
but the \{0110\} pattern for Ga, Ge, In, Sn, and Sb.

The final basis sets for most elements were then obtained simply by
adding the optimum (2d1f) exponents to the `triple-zeta' contraction
--- leading to a [3s3p2d1f] contracted basis set ---,
and the optimum (3d2f1g) exponents to the `quadruple-zeta' contraction
--- leading to a [4s4p3d2f1g] contracted basis set.
For Te and I, because of the peculiarities of the $d$ exponents noted
above, this procedure does not yield a satisfactory SDB-cc-pVTZ basis
set. By obtaining CISD natural orbitals for Te and I using $(3d1f)$
primitives, it was revealed that the highest-exponent primitive contributed
appreciably (and similarly) to the lowest two $d$-type natural orbitals, but that the 
latter are mainly distinguished by a sign change in the lowest-exponent 
$d$ primitive. Consequently, the two innermost $d$ primitives were 
contracted based on their coefficients in the lowest $d$ type natural orbital.
The slight added cost should be well outweighed by the greater reliability.
Considering the d-type ANOs in calculations with $(3d1f)$ primitives
on Sb, Sn, and In revealed that the same procedure might be beneficial for
In, but would not affect Sn or Sb. Therefore, in our final SDB-cc-pVTZ
basis sets, the $d$ functions in In, Te, and I are in fact $(3d)\rightarrow[2d]$
segmented contractions. 

The final basis sets generated are available on the Internet World Wide Web
at the Uniform Resource Locator 
\url{http://theochem.weizmann.ac.il/web/papers/SDB-cc.html}
in both Gaussian 98 and MOLPRO format.

\subsection{Diffuse function exponents}

For anionic systems and some very polar compounds, the availability of 
(diffuse-function) `augmented' basis sets, like the original aug-cc-pV$n$Z
basis sets\cite{avnz}, is essential. We have obtained diffuse functions for use
with our SDB-cc-pVTZ and SDB-cc-pVQZ basis sets using the following procedure:
(a) one low-exponent $s$ and $p$ function, each, were added to the $sp$ part of
the underlying basis
set and optimized simultaneously at the SCF level for the corresponding 
atomic anion; (b) successive angular momenta of the underlying basis
set were introduced, and one additional low-exponent primitive added
and optimized, at the CISD level for the corresponding atomic anion.
The final SDB-aug-cc-pVTZ and SDB-aug-cc-pVQZ basis sets are thus of
$[4s4p3d2f]$ and $[5s5p4d3f2g]$ quality, respectively.

\section{Application to diatomic molecules}

In order to validate our basis sets, we have carried out CCSD(T) 
calculations of the dissociation energy ($D_e$), bond length ($r_e$),
harmonic frequency ($\omega_e$) and first-order anharmonicity 
($\omega_ex_e$) of a number of third-row and fourth-row diatomic molecules
selected from the compilation by Huber and Herzberg\cite{Hub79}. 
CCSD(T) energies were computed at eleven points spaced evenly at 0.01 \AA\ 
intervals around the experimental $r_e$, a fifth-or sixth-order polynomial
in $r$ was fit, and a standard Dunham analysis\cite{Dun32} carried out 
on the resulting polynomial. For the open-shell systems and the constituent
atoms, the CCSD(T)
definition according to Ref.\cite{Wat93} was employed throughout.

Since we are using `large' cores, we also carried out calculations using
core polarization potentials (CPPs). For elements of groups IV, V, and VI the parameters
were taken from the work of Igel-Mann et al.\cite{Ige88}, although the cutoff
parameters given in that reference are not optimal for the ECP$nn$MWB pseudopotentials. For 
group III elements, optimal cutoffs were taken from Leininger et al.\cite{Lei97},
while optimal cutoffs for the halogens were taken from the online version
of the SDB pseudopotentials\cite{sddonline}. (The valence basis set was left
unchanged.)

For systems that include at most third-row atoms, all-electron calculations
could be carried out for comparison using 
the corresponding standard cc-pV$n$Z basis sets
\cite{Dun89,Woo93,Wil99}. $r_e$ and $\omega_e$ for these species are 
given in Table \ref{row3freq}, while $D_e$ values are given in Tables \ref{GaGeDe} and \ref{row3De}.
For the remaining diatomics (which include at least one fourth-row
atom), the corresponding data are found in Tables \ref{row4freq} and
\ref{row4De}, respectively.

In comparing such data with all-electron
calculations in which only valence electrons are correlated, it should be kept 
in mind that the CPPs approximately account for both inner-shell relaxation/polarization
("static core polarization") and inner-shell correlation
("dynamic core polarization"). Therefore, a direct comparison appears to be  somewhat
`unfair' to the all-electron calculations; on the other hand, since the standard
cc-pV$n$Z basis sets are by definition of minimal basis set quality in the
inner-shell orbitals, these basis sets are fairly limited in terms of flexibility
for static polarization. For heavier elements, it should also be kept in mind that 
the ECP calculations include relativistic effects at least approximately, while
their all-electron counterparts discussed here are entirely nonrelativistic.

For the late third-row species, it seems to be clear that the performance
of our SDB-cc-pV$n$Z basis sets is on a par with that of the all-electron
basis sets. Introduction of the core polarization potentials results in
a significant improvement in the computed bond distance: agreement
between SDB+CPP-cc-pVQZ and experimental bond lengths is particularly good
for many species. This conclusion is less clear for the harmonic frequencies,
where the known tendency\cite{ch} of CCSD(T) to slightly overestimate
harmonic frequencies may mask any small improvements. The computed
anharmonicities (not reported in Table \ref{row3freq}) agree very well 
between the various methods and experiment.

For the early third-row species, we noticed the at first sight peculiar
phenomenon (Table \ref{GaGefreq}) that, while our data with CPP are
in very good agreement with experiment, the all-electron bond lengths
are considerably too long, e.g. 0.05 \AA\ in GeF. (These differences
are too large to be plausibly ascribed to relativistic effects 
accounted for by the pseudopotentials.) The cause lies in the
impossibility to make a meaningful separation between `valence' and (3d) 
orbitals in these molecules: if correlation from the (3d) orbitals is
admitted, a dramatic improvement is seen in the computed bond distances.
Needless to say, such calculations are vastly more expensive than those
with the large-core pseudopotentials, and if the all-electron basis set would be 
expanded with the appropriate angular correlation functions for (3d)
correlation (i.e., high-exponent f and g functions), this would further
increase the cost differential. 

As an illustration, 
we will consider the GaH molecule in somewhat greater detail (Table \ref{tabGaH}).
The all-electron calculations with standard cc-pV$n$Z basis sets fortuitously 
reproduce $\omega_e$ very well, but overestimate the bond distance by almost 
0.03 \AA. In contrast, 4-electron ECP calculations with an 28-electron pseudopotential
both underestimate $\omega_e$ and overestimate $r_e$. 
Admitting (3d) correlation with the cc-pV$n$Z basis sets leads to a dramatic
shortening of $r_e$, but also to severely overestimated $\omega_e$ and 
anharmonicity. Obviously, this basis set needs to be significantly extended
before it is suitable for (3d) correlation. We have generated such basis
sets, denoted cc-pDVTZ and cc-pDVQZ, in the following manner.
All basis functions in the original cc-pVQZ basis set were retained, but
four additional $d$ primitives were decontracted. After this, successive
layers of $f$, $g$, and finally $h$ primitives were optimized at the
CISD level (13 electrons correlated) on top of the original basis set.
We found that the first h, second g, and third f function yielded similar
lowerings of the atomic energy, and hence added (3f2g1h) primitives to
the basis set.  (Exponents and other details can be found in the Supplementary
Material.) Then we restored the original $d$ functions and progressively
uncontracted primitives: while the first additional uncontracted $d$ yields
a very large energy lowering, the second adds a 10 millihartree amount comparable
to that of the $h$ functions, while lowerings decay rapidly after that.
Hence the final cc-pDVQZ basis set is of [7s6p6d5f3g1h] quality.
By similar arguments, we find that the cc-pVTZ basis set requires addition
of (2f1g) functions and decontraction of two additional $d$ primitives,
leading to a cc-pDVTZ basis set of [6s5p6d3f1g] quality. (While the $f$ exponents in the 
cc-pDVQZ basis set span a continuous range, a `gap' is present in the cc-pVTZ
case. A similar phenomenon is seen in an earlier Ga basis set of
Bauschlicher\cite{Bau98Ga}.)
These basis sets indeed do represent an improvement (Table \ref{tabGaH}) but
the 14-electron results clearly are still deficient in some respect. 
We considered also including (3s3p) correlation: to accommodate this,
we uncontracted two additional s and p primitives each in the cc-pDVTZ basis
set, as well as (to ensure adequate coverage of angular correlation from
these orbitals) one additional $d$ function. The resulting spectroscopic
constants are in excellent agreement with experiment, which might lead
to the conclusion that (3s3p) correlation is essential for a proper
description of GaH. However, as we reduce the number of correlated
electrons from 22 to 14, we see only quite minor effects on the 
spectroscopic constants. 
At that stage, the additional $d$ function
can be removed with essentially no effect on the computed spectroscopic
constants; the cc-pDVTZ+2s2p basis set is thus of [$8s7p6d3f1g$] quality. 
Correlating valence orbitals only leads to $r_e$ being too
long and $\omega_e$ too low, confirming that the excellent $\omega_e$
with standard cc-pV$n$Z basis sets is indeed the result of an error compensation.

We now consider the use of small-core ECPs. The effect of reoptimizing exponents
was deemed minimal: instead we simply (a) carried out an ECP10MWB Hartree-Fock
calculation with an uncontracted cc-pVTZ or cc-pVQZ basis set; (b) deleted
all primitives with coefficients$\times$degeneracies that are significantly
less than $10^{-4}$; repeated the SCF calculation and recontracted the
basis set with the orbital coefficients thus obtained. In the cc-pVTZ
basis set, we were able to delete the innermost (5s2p) primitives; in 
the cc-pVQZ basis set, the innermost (6s3p) primitives could be deleted.
The recontracted basis sets (which are of [4s4p5d3f1g] and [are again given in the supplementary material.
(As given, these basis sets are of [4s4p5d3f1g] and [5s5p6d5f3g1h] contracted size;
to this should be added the additional decontracted $s$ and $p$ primitives
mentioned above, leading to an SDB-cc-pDVTZ+2s2p basis set of
[$6s6p5d2f1g$] quality and an SDB-cc-pDVQZ basis set of 
[$8s8p6d5f3g1h]$ quality.)
We indeed find performance with these basis set-ECP combinations to be
quite satisfactory (Table \ref{tabGaH}).  (Note that for technical reasons, 
the SDB-cc-pDVQZ results do not include $h$ functions.)

It should also be noted that the effects of (3d) correlation, while
still important in accurate work, are significantly smaller with 
the cc-pDVTZ+2s2p and cc-pDVQZ+3s3p basis sets than with their 
less extended counterparts. The very large core correlation contributions
seen in such studies as Ref.\cite{Moc98} are thus at least in part
basis set artefacts.

Results for GeH (Table \ref{tabGaH}) follow similar trends as those
for GaH, although the deviation from experiment incurred by neglecting
(3d) correlation is definitely smaller. Continuing the series,
our computed results for AsH, SeH, and HBr suggest no need
for including (3d) correlation in these systems. 

We also applied the cc-pDVTZ and cc-pDVQZ basis sets to the polar Ga and Ge compounds
(Table \ref{GaGefreq}). A (sometimes notable) improvement is mainly seen in the
vibrational frequencies. Decontracting additional (sp) primitives in the 
Ga basis set was considered for GaF, and does not appear to greatly affect
results. This parallels a finding noted earlier\cite{cc} for inner-shell correlation
in first-row compounds, where flexibility of the core correlation basis set
appears to be more important for A--H than for A--B bonds. 

For  the fourth-row systems, only a comparison with experiment is possible.
Especially the SDB+CPP-pVQZ results agree very well with experiment, while
the errors for the SDB+CPP-pVTZ basis sets are not dissimilar from those
seen for the lighter-atom systems. A notable exception is constituted by 
a number of
indium compounds, for which abnormally short bond distances are found.
This problem has been noted previously for large-core pseudopotential
calculations on heavy group III halides\cite{sword}.
We attempted a number of calculations in which Bauschlicher's 
correlation-consistent basis set for In was used in conjunction with
regular cc-pV$n$Z basis sets on H--Ar and SDB-cc-pV$n$Z on Ga--Kr and
Sn--Xe. The In (4d) electrons were correlated in these calculations.
This completely resolves the problem. Discrepancies between all-electron
and ECP28MWB basis sets on In are not inconsistent with the expected 
magnitude of relativistic effects on $r_e$ and $\omega_e$. For InBr and
InI, consideration of a core polarization potential on the halogen has effects of 
-0.007 \AA\ and -0.013 \AA, respectively, on the bond distance,
bringing them into excellent agreement with experiment.
(Note that the +0.005 \AA\ discrepancy between computed and observed $r_e$(InCl)
found by Bauschlicher\cite{Bau99In} with his largest basis set appears to
be almost entirely due to $(2s2p)$ correlation in Cl: its inclusion reduces $r_e$
by 0.005 \AA. )

Finally, we shall consider dissociation energies. These are found
in Tables \ref{row3De}, \ref{GaGeDe}, and \ref{row4De}, together
with experimental data from two sources. These are the 1979
Huber and Herzberg (HH) book\cite{Hub79}, and a more recent compilation
by Kerr and Stocker (KS)\cite{De_CRC}  which contains
data through November 1998. 

All computed dissociation energies are corrected for atomic and 
molecular first-order spin-orbit splitting, with the data taken from
the experimental sources for the molecules and from 
Ref.\cite{atomsplit} for the atoms.

Aside from atomization energies with SDB-cc-pV$n$Z, SDB-aug-cc-pV$n$Z,
and all-electron cc-pV$n$Z basis sets, the tables contain extrapolations
to the infinite basis limit using the expression taken from W1 theory\cite{w1}:
$E_\infty=E[VQZ]+(E[VQZ]-E[VTZ])/((4/3)^{3.22}-1)$, where the exponent 
3.22 is specific to the VTZ/VQZ basis set combination. This is in fact
a damped variant of the simple $A+B/l^3$ formula of Halkier et al.\cite{Hal98}:
the damping is required\cite{w1} because the VTZ and VQZ basis sets are still not extended
enough and lead to overshooting if the $A+B/l^3$ formula is applied to them.
(The latter is the extrapolation of choice for larger basis sets.)

One conspicuous feature of the experimental results is just how uncertain
they are for many molecules in these tables. For the late third-row
systems, agreement between experiment and
our extrapolated results including CPP
is excellent for those molecules where the experimental
value is precisely known. For most of the other systems, the computed value
falls within the combined uncertainties of the experimental values.
Agreement in fact appears to be slightly better than for the all-electron
calculations, but this is not an entirely `fair' comparison since the
latter include neither inner-shell correlation nor scalar relativistic
corrections, while both are included approximately in the SDB+CPP results
through the core-polarization potential and the relativistic pseudopotential,
respectively. 

For the Ga, Ge, and In compounds, experimental dissociation energies are so uncertain
that a meaningful comparison is essentially impossible. For those fourth-row
systems where precise experimental data are available, agreement with 
experiment is still quite satisfactory, albeit less good than for the
third-row compounds. In particular, an account for higher-order spin-orbit
effects might be mandatory for some of the iodine compounds. Dolg\cite{Dol96}
carried out benchmark calculations on the hydrogen halides and dihalides, and
found near-exact spin-orbit contributions to $D_e$(HI) and $D_e$(I$_2$) of
0.26 and 0.49 eV, respectively: simply considering the fine structures of
the constituent atoms (as done here) yields 0.315 and 0.63 eV, respectively.
In other words, our calculated $D_e$ values for HI and I$_2$ are intrinsically
too low by 0.07 and 0.14 eV, respectively.

As expected, the use of (diffuse function)
`augmented' basis sets yields improved results for
$r_e$ and $\omega_e$ of highly polar molecules (e.g., GaF); for $D_e$ values,
differences of up to 0.05 eV are seen after extrapolation, which are definitely
significant in accurate thermochemical work. As is
the general rule\cite{l4,w1}, the addition of diffuse functions
considerably improves the success
of extrapolation methods and improves agreement with (precise)
experimental dissociation energies.

Finally, we should address the question whether or not the RECPs used here
provide an approximate account for scalar relativistic effects.
Visscher and coworkers studied relativistic effects on the hydrogen halogenides\cite{Vis1},
dihalogenides\cite{Vis2}, and interhalogenides\cite{Vis3} by means of full four-component
relativistic CCSD(T) as implemented by Visscher, Lee, and Dyall\cite{relccsd}.
(Pisani and Clementi\cite{Pis95} also carried out Dirac-Fock calculations on the chalcogen 
hydrides --- including SeH --- and found an effect of -0.005 \AA\ on $r_e$.)
Since down to Br, the effects are fairly small
(e.g. +0.003 \AA\ and --6 cm$^{-1}$ in BrF), a comparison between all-electron and ECP
results is somewhat dubious as an indicator for the recovery of relativistic effects.
Given however the sizable relativistic contributions found in that work for the
iodine compounds (e.g. IF: +0.012 \AA\ and -23 cm$^{-1}$), the level of agreement with
experiment found in the present paper is somewhat hard to explain 
unless the ECPs indeed 
recover most of the scalar relativistic effects.

\section{Conclusions}

We have derived (fairly) compact valence basis sets of cc-pV$n$Z and
aug-cc-pV$n$Z quality ($n$=T,Q) for the elements Ga--Kr and In--Xe,
to be used in conjunction with large-core Stuttgart-Dresden-Bonn
pseudopotentials. For the third row, the basis sets appear to be
quite comparable to the corresponding all-electron cc-pV$n$Z basis
sets. Agreement with experiment is quite satisfactory for compounds 
of the later heavy p-block elements. Highly accurate calculations 
on Ga and, to a lesser extent, Ge compounds require treating the (3d)
electrons explicitly: we propose $(3d)$-correlation basis
sets for these elements. For In compounds, inclusion of (4d) correlation
is a must, as previously found by Bauschlicher\cite{Bau99In}: we recommend
the basis sets in that reference.

Our principal objective was having extended basis sets available for 
studies on organometallic compounds, including those with one or more
heavy group V, VI, and VII elements. This objective appears to have
been reached. 

\acknowledgments

JM is the incumbent of the Helen and Milton A. Kimmelman Career 
Development Chair.
Research at the Weizmann Institute 
was supported by the Minerva Foundation, Munich, Germany, and
by the {\em Tashtiyot} program of the Ministry of Science (Israel).
AS acknowledges a Minerva Postdoctoral Fellowship. 
The authors
would like to thank Dr. Charles W. Bauschlicher Jr. (NASA Ames
Research Center) for critical
reading of the manuscript prior to submission.

\section*{Supplementary material}

The SDB-cc-pVTZ, SDB-cc-pVQZ, cc-pDVTZ, and cc-pDVQZ 
basis sets developed in this paper are available for download
on the Internet World Wide Web at the URL
\url{http://theochem.weizmann.ac.il/web/papers/SDB-cc.html}

\section*{Appendix: $f$-and $g$-function exponents for the transition metals}

For use in conjunction with the above SDB-cc-pVTZ basis set on Ga-Kr
and In--Xe, and the standard cc-pVTZ basis set on the first two
rows of the periodic table, we recommend the following basis
set/ECP combination for transition metals.

For first-row transition metals, the pseudopotential denoted as ECP10MDF\cite{Dol87} (which has a small 10-electron core)
was used in conjunction with the [6s5p3d] contraction of an (8s7p6d) primitive set given in Ref.\cite{Dol87}. For
second-and third-row transition metals, we used the ECP28MWB and ECP60MWB quasirelativistic pseudopotentials,
respectively, as given in Ref.\cite{And90}, together with the [6s5p3d] contracted valence basis sets given in
the same reference.

Two $(4f)$-type functions and one $(5g)$-type function were added, and their exponents optimized at the CISD
level for the lowest-lying $(s)^1(d)^{n-1}$ and $(s)^2(d)^{n-2}$ states. (In addition, optimizations were
carried out for the $(s)^0(d)^{10}$ ground state of Pd.) Proper symmetry and spin eigenfunctions
were used for the Hartree-Fock reference, and only valence electrons were correlated. The optimum exponents
for the two states considered (three in the case of Pd) are not very different: we recommend their averages
as the $f$ and $g$ exponents, which are given in Table \ref{transition}.

\begin{table}
\caption{CCSD(T) and experimental spectroscopic constants ($R_e$ in \AA, $\omega_e$ in cm$^{-1}$) for molecules containing third row atoms\label{row3freq}}
\squeezetable
\begin{tabular}{lc*{8}{d}} 
molecule& basis & \multicolumn{4}{c}{$R_e$} &  \multicolumn{4}{c}{$\omega_e$} \\
        & & SDB & CPP & all $e^-$ & Exp. 
          & SDB & CPP & all $e^-$ & Exp. \\
\hline
AlBr & VTZ   & 2.3149 & 2.3088 & 2.3271 & 2.294807 & 383.0  & 384.4  & 377.3  & 378.0  \\
     & VQZ   & 2.3073 & 2.3014 & 2.3180 &          & 381.1  & 382.3  & 377.0  &        \\
     & AVTZ  &        & 2.3164 &        &          &        & 377.9  &        &        \\
     & AVQZ  &        & 2.3050 &        &          &        & 374.6  &        &        \\
As$_2$&VTZ   & 2.1265 & 2.1126 & 2.1348 & 2.1026   & 424.0  & 426.8  & 424.6  & 429.55 \\
      &VQZ   & 2.1098 & 2.0963 & 2.1284 &          & 432.1  & 435.1  & 428.7  &        \\
AsF $^3\Sigma^-$ 
     & VTZ   & 1.7298 & 1.7224 & 1.7464 & 1.7360   & 697.0  & 700.3  & 698.1  & 685.78  \\
     & VQZ   & 1.7291 & 1.7217 & 1.7446 &          & 693.6  & 696.9  & 697.1  &         \\
     & AVTZ  &        & 1.7298 &        &          &        & 687.1  &        &        \\
     & AVQZ  &        & 1.7185 &        &          &        & 700.9  &        &        \\
AsH $^3\Sigma^-$
     & VTZ   & 1.5320 & 1.5269 & 1.5354 & 1.5231$^a$ & 2135.3 & 2138.2 & 2151.9 & 2155.503$^a$\\
     & VQZ   & 1.5286 & 1.5234 & 1.5335 &          & 2156.0 & 2159.5 & 2161.1 &         \\
AsN  & VTZ   & 1.6292 & 1.6227 & 1.6374 & 1.61843   & 1063.2 & 1069.7 & 1064.5 & 1068.54 \\
     & VQZ   & 1.6188 & 1.6124 & 1.6326 &          & 1079.1 & 1085.8 & 1071.9 &         \\
AsO $^2\Pi$   
     & VTZ   & 1.6260 & 1.6200 & 1.6378 & 1.6236   & 973.6  & 978.2  & 971.2  & 967.08  \\
     & VQZ   & 1.6199 & 1.6138 & 1.6344 &          & 978.5  & 983.2  & 975.5  &         \\
AsP  & VTZ   & 2.0216 & 2.0142 & 2.0276 & 1.999    & 597.7  & 600.2  & 597.3  & 604.02  \\
     & VQZ   & 2.0083 & 2.0011 & 2.0194 &          & 607.8  & 610.3  & 604.1  &         \\
AsS $^2\Pi$ 
     & VTZ   & 2.0393 & 2.0319 & 2.0435 & 2.0174   & 559.5  & 561.5  & 561.3  & 567.94  \\
     & VQZ   & 2.0246 & 2.0174 & 2.0349 &          & 570.0  & 571.9  & 568.0  &         \\
BBr  & VTZ   & 1.8964 & 1.8908 & 1.9060 & 1.8882   & 690.5  & 693.8  & 683.7  & 684.31  \\
     & VQZ   & 1.8942 & 1.8887 & 1.9034 &          & 685.8  & 688.6  & 681.1  &         \\
     & AVTZ  &        & 1.8918 & 1.9068 &          &        & 684.4  & 677.9  &        \\
     & AVQZ  &        & 1.8894 & 1.9044 &          &        & 700.9  & 678.8  &        \\
Br$_2$&VTZ   & 2.3138 & 2.3014 & 2.3108 & 2.28105  & 313.6  & 316.4  & 319.3  & 325.321 \\
     & VQZ   & 2.2970 & 2.2856 & 2.2983 &          & 323.8  & 326.1  & 325.9  &         \\
     & AVTZ  &        & 2.2941 & 2.3127 &          &        & 318.1  & 317.3  &        \\
     & AVQZ  &        & 2.2808 & 2.2986 &          &        & 326.9  & 325.1  &        \\
BrCl & VTZ   & 2.1616 & 2.1555 & 2.1627 & 2.136065 & 434.3  & 436.2  & 435.9  & 444.276 \\
     & VQZ   & 2.1491 & 2.1432 & 2.1504 &          & 442.2  & 443.9  & 443.3  &         \\
     & AVTZ  &        & 2.1545 &        &          &        & 434.6  &        &        \\
     & AVQZ  &        & 2.1420 &        &          &        & 443.5  &        &        \\
BrF  & VTZ   & 1.7680 & 1.7628 & 1.7685 & 1.75894  & 661.0  & 664.0  & 666.5  & 670.75  \\
     & VQZ   & 1.7611 & 1.7559 & 1.7619 &          & 672.0  & 674.5  & 677.0  &         \\
     & AVTZ  &        & 1.7622 &        &          &        & 667.7  &        &        \\
     & AVQZ  &        & 1.7547 &        &          &        & 676.5  &        &        \\
CSe  & VTZ   & 1.6921 & 1.6863 & 1.6966 & 1.67647  & 1019.3 & 1025.4 & 1024.1 & 1035.36 \\
     & VQZ   & 1.6834 & 1.6776 & 1.6911 &          & 1032.7 & 1038.8 & 1031.2 &         \\
HBr  & VTZ   & 1.4147 & 1.4106 & 1.4203 & 1.414435 & 2656.1 & 2665.0 & 2660.0 & 2648.975\\
     & VQZ   & 1.4155 & 1.4112 & 1.4205 &          & 2657.7 & 2667.2 & 2661.2 &         \\
     & AVTZ  & 1.4166 & 1.4124 & 1.4213 &          & 2647.2 & 2656.2 & 2657.2 &         \\
     & AVQZ  & 1.4163 & 1.4120 & 1.4211 &          & 2651.9 & 2661.5 & 2658.1 &         \\
SeH  & VTZ   & 1.4702 & 1.4654 & 1.4731 & 1.46432(6)$^b$& 2398.2 & 2406.3 & 2417.6 & 2421.715(23)$^b$\\
     & VQZ   & 1.4680 & 1.4630 & 1.4721 &               & 2416.6 & 2425.5 & 2425.4 \\
NSe $^2\Pi$    
     & VTZ   & 1.6659 & 1.6606 & 1.6671 & 1.6518   & 946.8  & 952.3  & 955.8  & 956.81  \\
     & VQZ   & 1.6539 & 1.6487 & 1.6599 &          & 964.9  & 970.4  & 966.1  &         \\
Se$_2$ $^3\Sigma^-_g$
     & VTZ   & 2.1911 & 2.1783 & 2.1915 & 2.1660   & 381.3  & 384.6  & 384.7  & 385.303 \\
     & VQZ   & 2.1752 & 2.1628 & 2.1830 &          & 388.3  & 391.5  & 389.6  &         \\
SeO $^3\Sigma^-$
     & VTZ   & 1.6476 & 1.6425 & 1.6500 & 1.6484   & 914.4  & 918.8  & 922.2  & 914.69  \\
     & VQZ   & 1.6379 & 1.6328 & 1.6454 &          & 924.9  & 929.3  & 926.6  &         \\
SeS $^3\Sigma^-$
     & VTZ   & 2.0546 & 2.0482 & 2.0558 & 2.0367   & 547.7  & 550.2  & 549.5  & 555.56  \\
     & VQZ   & 2.0407 & 2.0344 & 2.0458 &          & 556.2  & 558.7  & 556.6  &         \\
SiSe & VTZ   & 2.0834 & 2.0770 & 2.0882 & 2.058324 & 572.0  & 574.6  & 571.3  & 580.0   \\
     & VQZ   & 2.0713 & 2.0650 & 2.0790 &          & 578.5  & 581.2  & 576.9  &         \\
     & AVTZ  & 2.0840 & 2.0775 & 2.0887 &          & 569.3  & 572.1  & 569.1  &         \\
     & AVQZ  & 2.0723 & 2.0659 & 2.0794 &          & 576.1  & 578.9  & 576.1  &         \\
\end{tabular}
SDB: calculations using large-core SDB pseudopotentials; CPP: ditto with core polarization potentials added.
$^a$ AsH ($R_e$ and $\omega_e$): K. D. Hensel, R. A. Hughes, and J. M. Brown \jcsfii{91}{2999}{1995}.
$^b$ SeH ($R_e$ and $\omega_e$): R. S. Ram and P. F. Bernath, \jmsp{203}{9}{2000}.
\end{table}

\begin{table}
\caption{CCSD(T) and experimental spectroscopic constants ($R_e$ in \AA, $\omega_e$ in cm$^{-1}$) for molecules containing fourth row atoms\label{row4freq}}
\squeezetable
\begin{tabular}{lc*{6}{d}} 
molecule& basis & \multicolumn{3}{c}{$R_e$} &  \multicolumn{3}{c}{$\omega_e$} \\
        & & SDB & CPP & Exp. 
          & SDB & CPP & Exp. \\
\hline
AlI  & VTZ   & 2.5591 & 2.5478 & 2.537102 & 321.1  & 323.1  & 316.1 \\
     & VQZ   & 2.5521 & 2.5408 &          & 318.4  & 320.1  &       \\
     & AVTZ  &        & 2.5519 &          &        & 316.3  &       \\
     & AVQZ  &        & 2.5441 &          &        & 316.8  &       \\ 
GaI  & VTZ   & 2.6240 & 2.5955 & 2.57467  & 212.2  & 213.6  & 216.6  \\
     & VQZ   & 2.6150 & 2.5871 &          & 210.1  & 211.3  &        \\
GeTe & VTZ   & 2.3802 & 2.3596 & 2.340165 & 313.1  & 315.7  & 323.9  \\
     & VQZ   & 2.3554 & 2.3358 &          & 312.4  & 322.6  &        \\
HI   & VTZ   & 1.6147 & 1.6073 & 1.60916  & 2314.7 & 2325.7 & 2309.014\\
     & VQZ   & 1.6131 & 1.6054 &          & 2320.3 & 2332.7 &         \\
     & AVTZ  & 1.6162 & 1.6088 &          & 2310.4 & 2321.4 &         \\
     & AVQZ  & 1.6135 & 1.6058 &          & 2318.1 & 2330.3 &         \\
I$_2$& VTZ   & 2.7072 & 2.6831 & 2.6663   & 212.2  & 215.8  & 214.502 \\
     & VQZ   & 2.6876 & 2.6655 &          & 218.4  & 221.3  &         \\
IBr  & VTZ   & 2.5049 & 2.4870 & 2.468989 & 263.6  & 266.9  & 268.640 \\
     & VQZ   & 2.4862 & 2.4698 &          & 271.5  & 274.2  &         \\
     & AVTZ  &        & 2.4805 &          &        & 267.9  &       \\
     & AVQZ  &        & 2.4652 &          &        & 274.6  &       \\ 
ICl  & VTZ   & 2.3482 & 2.3371 & 2.320878 & 383.2  & 386.1  & 384.293 \\
     & VQZ   & 2.3349 & 2.3239 &          & 388.3  & 391.0  &         \\
     & AVTZ  &        & 2.3358 &          &        & 385.4  &       \\
     & AVQZ  &        & 2.3225 &          &        & 391.6  &       \\ 
IF   & VTZ   & 1.9296 & 1.9204 & 1.90975  & 611.8  & 616.2  & 610.24  \\
     & VQZ   & 1.9158 & 1.9066 &          & 622.0  & 626.1  &         \\
     & AVTZ  &        & 1.9196 &          &        & 616.8  &       \\
     & AVQZ  &        & 1.9070 &          &        & 625.7  &       \\ 
InBr & VTZ   & 2.5377 & 2.5029 & 2.54318  & 229.1  & 231.2  & 221.0   \\
     & VQZ   & 2.5080 & 2.4716 &          & 228.7  & 230.5  &         \\
     & CWB(t)& 2.5556 &        &          & 224.0\\ 
     & CWB(q)& 2.5486 &        &          & 224.0\\
     &ACWB(t)& 2.5677 & 2.5606 &          & 218.1 & 219.0\\
     &ACWB(q)& 2.5540 & 2.5474 &          & 220.7 & 221.6\\
InCl & VTZ   & 2.3849 & 2.3553 & 2.401169 & 327.1  & 330.0  & 317.4   \\
     & VQZ   & 2.3571 & 2.3246 &          & 326.7  & 328.6  &         \\
&ACWB(t)\cite{Bau99In}& 2.423 &       &   & 309    &     \\
&ACWB(q)\cite{Bau99In}& 2.411 &       &   & 314    &     \\
&ACWB(5)\cite{Bau99In}& 2.406 &       &   & 316\\
all e-&CWB(t)\cite{Bau99In}&2.423 &   &   & 317\\
all e-&CWB(q)\cite{Bau99In}&2.412 & 2.4068$^e$ &   & 318 & 316.0$^e$\\
all e-&CWB(5)\cite{Bau99In}&2.406 &   &   & 319\\
& (a)  & 2.4206 &        &          & 319.4  &        &        \\
InF  & VTZ   & 1.9329 & 1.9109 & 1.985396 & 568.8  & 577.0  & 535.35  \\
     & VQZ   & 1.8976 & 1.8729 &          & 579.4  & 587.1  &         \\
     & CWB(t)& 1.9833 &        &          & 538.5  &        \\
     & CWB(q)& 1.9822 &        &          & 541.3  &        \\
     &ACWB(t)& 1.9910 &        &          & 526.6  &        \\
     &ACWB(q)& 1.9853 &        &          & 534.3  &        \\
all e-&CWB(t)& 1.9833 &        &          & 550.7  \\
all e-&CWB(q)& 1.9821 &        &          & 553.7  \\
all e-&ACWB(t)&1.9910 &        &          & 539.4 \\
all e-&ACWB(q)&1.9857 &        &          & 546.6 \\
InH  & VTZ   & 1.8653 & 1.8517 & 1.8380   & 1482.3 & 1472.7 & 1476.04 \\
     & VQZ   & 1.8317 & 1.8171 &          & 1503.5 & 1497.3 &         \\
     & CWB(t)& 1.8395 &        &          & 1462.4 &        \\
     & CWB(q)& 1.8374 &        &          & 1471.9 &       \\
all e-&CWB(t)& 1.8491 &        &          & 1493.4 \\
all e-&CWB(q)& 1.8473 &        &          & 1504.4 \\
InI  & VTZ   & 2.7698 & 2.7273 & 2.75365  & 179.0  & 181.1  & 177.1   \\
     & VQZ   & 2.7412 & 2.6983 &          & 178.9  & 180.6  &         \\
     & CWB(t)& 2.7740 & 2.7605 &          & 177.8  & 179.0 \\
     & CWB(q)& 2.7674 & 2.7543 &          & 177.0  & 178.1 \\
Sb$_2$&VTZ   & 2.5294 & 2.5045 & 2.476$^b$& 266.3  & 268.2  & 269.623$^b$ \\
     & VQZ   & 2.5056 & 2.4816 &          & 273.0  & 275.1  &         \\
SbF $^3\Sigma^-$ 
     & VTZ   & 1.9252 & 1.9138 & 1.9177   & 615.3  & 619.8  & 609.0$^d$\\
     & VQZ   & 1.9137 & 1.9023 &          & 620.0  & 623.9  &         \\
SbH $^3\Sigma^-$ 
     & VTZ   & 1.7259 & 1.7181 & 1.7107$^c$& 1910.9 & 1914.0 & 1923.179 $^c$\\
     & VQZ   & 1.7188 & 1.7107 &          & 1930.3 & 1933.6 &         \\
SbP  & VTZ   & 2.2335 & 2.2211 & 2.205    & 495.0  & 497.8  & 500.07  \\
     & VQZ   & 2.2183 & 2.2058 &          & 503.5  & 506.5  &         \\
SnO  & VTZ   & 1.8426 & 1.8297 & 1.832505 & 763.3  & 768.7  & 814.6   \\
     & VQZ   & 1.8271 & 1.8137 &          & 781.9  & 788.5  &         \\
SnS  & VTZ   & 2.2382 & 2.2220 & 2.209026 & 466.9  & 468.6  & 487.26  \\
     & VQZ   & 2.2214 & 2.2050 &          & 474.9  & 476.8  &         \\
SnSe & VTZ   & 2.3595 & 2.3379 & 2.325601 & 316.3  & 318.1  & 331.2   \\
     & VQZ   & 2.3398 & 2.3181 &          & 321.0  & 323.2  &         \\
     & AVTZ  & 2.3541 & 2.3319 &          & 315.5  & 317.9  &         \\
     & AVQZ  & 2.3421 & 2.3202 &          & 319.2  & 321.6  &         \\
SnTe & VTZ   & 2.5563 & 2.5294 & 2.522814 & 253.5  & 255.4  & 259.5   \\
     & VQZ   & 2.5359 & 2.5091 &          & 256.9  & 258.9  &         \\
     & AVTZ  & 2.5492 & 2.5218 &          & 253.3  & 255.6  &         \\
     & AVQZ  & 2.5375 & 2.5104 &          & 255.6  & 257.8  &         \\
\end{tabular}
(a) using ECP28MDF for In, (4d) electrons correlated \\
(b)  Sb$_2$ ($R_e$, $\omega_e$): H. Sontag and R. Weber, \jmsp{91}{72}{1982}.\\
(c)  SbH ($R_e$, $\omega_e$): R.-D. Urban, K. Essig, and H. Jones, \jcp{99}{1591}{1993}.\\
(d)  SbF ($\omega_e$): D. K. W. Wang, W. E. Jones, F. Pr{\'e}vot, and R. Colin, \jmsp{49}{377}{1974}.\\
(e)  This work, using the MTavqz basis set\cite{w1} on Cl and including (2s,2p) correlation in Cl.\\
CWB(t) and CWB(q) indicate the Bauschlicher\cite{Bau99In} cc-pVTZ and cc-pVQZ basis sets on indium,
and regular cc-pV$n$Z or SDB-ccpV$n$Z basis sets on the other atom.
ACWB(t) and ACWB(q) indicate the same, but in conjunction with an augmented basis set on the other atom.\\
\end{table}

\begin{table}
\caption{CCSD(T) and experimental spectroscopic constants ($R_e$ in \AA, $\omega_e$ in cm$^{-1}$) for diatomics involving Ga or Ge and an electronegative
element\label{GaGefreq}}
\squeezetable
\begin{tabular}{lc*{4}{d}l*{5}{d}} 
molecule& & \multicolumn{4}{c}{$R_e$} &  \multicolumn{4}{c}{$\omega_e$} \\
        & & SDB & CPP & all $e^-$ (a)  & all $e^-$ (b)  & Exp. 
          & SDB & CPP & all $e^-$ (a)  & all $e^-$ (b)  & Exp. \\
\hline
GaBr&VTZ  & 2.3912 & 2.3681 & 2.4013 & 2.3618 & 2.35248  & 261.7  & 262.6  & 268.6  & 274.9  & 263.0    \\        
    &VQZ  & 2.3829 & 2.3602 & 2.4007 & 2.3644 &          & 258.7  & 259.5  & 266.6  & 266.8  &        \\
13$e^-$&DVTZ & 2.3564 &        &        & 2.3668 &          & 275.5  &        &        & 270.9  \\
    &DVQZ & 2.3486 &        &        & 2.3656 &          & 268.9  &        &        & 267.7  \\
GaCl&VTZ  & 2.2411 & 2.2227 & 2.2554 & 2.2031 & 2.201690 & 353.5  & 353.6  & 381.3  & 382.4  & 365.3  \\
    &VQZ  & 2.2320 & 2.2135 & 2.2572 & 2.2010 &          & 350.7  & 351.1  & 373.3  & 370.4  &        \\
13$e^-$&DVTZ & 2.2063 &        &        & 2.2070 &          & 372.9  &        &        & 374.7  \\
    &DVQZ & 2.1987 &        &        & 2.2044 &          & 368.4  &        &        & 370.1  \\
GaF &VTZ  & 1.7851 & 1.7688 & 1.8031 & 1.7756 & 1.774369 & 589.0  & 589.8  & 631.9  & 652.3  & 622.2  \\
    &VQZ  & 1.7753 & 1.7587 & 1.8043 & 1.7697 &          & 586.3  & 588.0  & 625.4  & 644.8  &        \\
13$e^-$&DVTZ & 1.7704 &        &        & 1.7709 &          & 638.7  &        &        & 644.2  \\ 
    &DVQZ & 1.7674 &        &        & 1.7716 &          & 636.1  &        &        & 637.6  \\
    &ADVTZ& 1.7798 &        &        & 1.7812 &          & 619.7  &        &        & 623.7  \\
    &ADVQZ& 1.7693 &        &        & 1.7741 &          & 631.2  &        &        & 628.9  \\
+2s2p&DVTZ& 1.7711 &        &        &        &          & 629.1  &        &        &        \\
+3s3p&DVQZ& 1.7714 &        &        &        &          & 629.3  &        &        &        \\
GeF $^2\Pi$
    &VTZ  & 1.7655 & 1.7550 & 1.8047 & 1.7469 & 1.7452   & 642.6  & 645.3  & 746.2  & 688.6  & 665.67 \\
    &VQZ  & 1.7577 & 1.7473 & 1.8096 & 1.7424 &          & 640.3  & 642.8  & 746.1  & 681.2  &        \\
13$e^-$&DVTZ & 1.7411 &        &        & 1.7450 &          & 683.1  &        &        & 677.6 \\
13$e^-$&DVQZ & 1.7407 &        &        & 1.7424 &          & 675.3  &        &        & 679.5 \\
GeO &VTZ  & 1.6382 & 1.6290 & 1.6491 & 1.6341 & 1.624648 & 925.1  & 930.8  & 987.4  & 993.7  & 985.5  \\
    &VQZ  & 1.6295 & 1.6205 & 1.6485 & 1.6276 &          & 943.2  & 949.1  & 989.3  & 1000.5 &        \\
13$e^-$&DVTZ & 1.6254 &        &        & 1.6293 &          & 995.4  &        &        & 992.3\\
13$e^-$&DVQZ & 1.6214 &        &        & 1.6269 &          & 996.7  &        &        & 996.6\\
GeS &VTZ  & 2.0461 & 2.0356 & 2.0416 & 2.0280 & 2.012086 & 552.2  & 554.1  & 568.4  & 578.0  & 575.8  \\
    &VQZ  & 2.0300 & 2.0200 & 2.0356 & 2.0192 &          & 560.5  & 562.4  & 572.3  & 578.2  &        \\
13$e^-$&DVTZ & 2.0207 &        &        & 2.0284 &          & 577.9  &        &        & 572.3 \\
13$e^-$&DVQZ & 2.0133 &        &        & 2.0190 &          & 577.5  &        &        & 578.2 \\
GeSe&VTZ  & 2.1743 & 2.1580 & 2.1713 & 2.1574 & 2.134629 & 384.8  & 387.4  & 397.5  & 403.7  & 408.7  \\
    &VQZ  & 2.1531 & 2.1377 & 2.1658 & 2.1506 &          & 393.5  & 444.5  & 400.2  &403.5  &        \\
13$e^-$&DVTZ & 2.1479 &        &        & 2.1583 &          & 403.2  &        &        &401.4 \\
13$e^-$&DVQZ & 2.1370 &        &        & 2.1495 &          & 405.0  &        &        &403.8 \\
\end{tabular}
(a) all d electrons of Ga, Ge are frozen for the all electron calculation.\\
(b) d-shell correlated.\\
\end{table}

\begin{table}
\caption{Spectroscopic constants for GaH and GeH.\label{tabGaH} All results at the CCSD(T) level.}
\squeezetable
\begin{tabular}{lllcccccc}
Nonvalence       & Pseudopotential & Basis set & $r_e$ & $\omega_e$ & $\omega_ex_e$ & $r_e$ & $\omega_e$ & $\omega_ex_e$ \\
$e^-$ correlated &                 &           & \AA   & cm$^{-1}$  & cm$^{-1}$ & \AA   & cm$^{-1}$  & cm$^{-1}$\\
\hline
&&&\multicolumn{3}{c}{GaH}&\multicolumn{3}{c}{GeH}\\
\multicolumn{3}{c}{Experiment} & 1.660149(2)$^a$ & 1603.9559(20)$^b$ & 28.4227$^b$ & 1.58724$^c$ & 1900.3820$^c$ & 33.5024(28)\\
--- & all-electron & cc-pVDZ  & 1.6860 & 1610.0 & 26.43 & 1.6083 & 1896.1 & 32.96\\
--- & all-electron & ditto full CI &1.6867 & 1607.0 & 26.55 & 1.6090 & 1891.7 & 33.24\\
--- & all-electron & cc-pVTZ  & 1.6878 & 1603.2 & 26.80 & 1.6039 & 1897.9 & 32.80 \\
--- & all-electron & cc-pVQZ  & 1.6864 & 1604.3 & 27.74 & 1.6021 & 1903.1 & 33.14 \\
(3d)& all-electron & cc-pVTZ & 1.6573 & 1691.8 & 31.15 & 1.5860 & 1956.9 & 36.24\\
(3d)& all-electron & cc-pVQZ & 1.6449 & 1698.5 & 40.56 & 1.5764 & 1965.0 & 43.78\\
--- & ECP28MWB & SDB-cc-pVTZ  & 1.6969 & 1530.9 & 24.27 & 1.6084 & 1844.4 & 30.67\\
--- &          &  + CPP       & 1.6876 & 1524.2 & 24.35 & 1.6014 & 1845.8 & 30.72\\
--- & ECP28MWB & SDB-cc-pVQZ  & 1.6820 & 1546.3 & 25.09 & 1.5990 & 1862.3 & 31.30\\
--- &          &  + CPP       & 1.6729 & 1540.4 & 25.13 & 1.5921 & 1863.5 & 31.46\\
(3d)& all-electron & cc-pDVTZ & 1.6582 & 1658.2 & 32.91 & 1.5893 & 1900.4 & 31.65\\
(3d)& all-electron & cc-pDVQZ & 1.6565 & 1653.6 & 36.20 & 1.5821 & 1950.2 & 42.03\\
(3d)& all-electron & cc-pDVTZ+2d     & 1.6500 & 1659.5 & 33.07 \\
(3s,3p,3d)& all-electron & cc-pDVTZ+2s2p1d & 1.6647 & 1602.5 & 26.56 \\
(3d)& all-electron & cc-pDVTZ+2s2p1d & 1.6635 & 1604.4 & 25.82 \\
--- & all-electron & cc-pDVTZ+2s2p1d & 1.6863 & 1598.0 & 25.56 \\
(3d)& all-electron & cc-pDVTZ+2s2p   & 1.6637 & 1604.0 & 25.69 & 1.5889 & 1906.2 & 32.30\\
--- & all-electron & cc-pDVTZ+2s2p   & 1.6863 & 1597.6 & 25.75 & 1.6020 & 1899.3 & 32.64\\
(3d)& ECP10MWB     & cc-pDVTZ        & 1.6550 & 1649.3 & 31.47 & 1.5783 & 1950.6 & 37.24\\
(3d)& ECP10MWB     &  cc-pDVQZ       & 1.6446 & 1662.8 & 37.57 & 1.5751 & 1941.7 & 40.79\\
(3d)& ECP10MWB     & cc-pDVTZ+2s2p   & 1.6613 & 1588.8 & 25.09 & 1.5865 & 1893.2 & 32.12\\
--- & ECP10MWB     & cc-pDVTZ+2s2p   & 1.6838 & 1584.4 & 25.12 & 1.5993 & 1887.7 & 32.40\\
(3d)& ECP10MWB     & cc-pDVQZ+2s2p   & 1.6584 & 1607.4 & 27.33 \\
(3d)& ECP10MWB     & cc-pDVQZ+3s3p   & 1.6586 & 1605.6 & 26.92 & 1.5881 & 1910.8 & 32.86\\
--- & ECP10MWB     &cc-pDVQZ+3s3p   &  1.6829 & 1592.1 & 27.38 & 1.6006 & 1897.4 & 32.40\\
(3s,3p,3d)& ECP10MWB&cc-pDVQZ+3s3p   & 1.6602 & 1601.5 & 26.00 & 1.5898 & 1907.4 & 32.82\\
\end{tabular}
$^a$ M. Molski, \jmsp{182}{1}{1997}.\\
$^b$ F. Ito, T. Nakanago, H. Jones, \jmsp{164}{379}{1994}.\\
$^c$ J. P. Towle and J. M. Brown, \mph{78}{249}{1993}.\\
\end{table}

\begin{table}
\caption{Binding energies ($D_e$ in eV) for molecules containing third row atoms. \label{row3De}}
\squeezetable
\begin{tabular}{lc*{11}{c}}
molecule& \multicolumn{3}{c}{SDB} & \multicolumn{3}{c}{CPP} & \multicolumn{3}{c}{all $e^-$} & \multicolumn{2}{c}{Experiment} \\
        & VTZ & VQZ & $\infty$ & VTZ & VQZ & $\infty$ & VTZ & VQZ & $\infty$ &  HH\cite{Hub79} & KS\cite{De_CRC}\\ 
\hline
AlBr                  & 4.31 & 4.45 & 4.55 & 4.33 & 4.48 & 4.57 & 4.29 & 4.43 & 4.52 & 4.43 & 4.42$\pm$0.06\\
ditto aug-cc          &      &      &      & 4.35 & 4.49 & 4.58 \\
As$_2$                & 3.42 & 3.78 & 4.02 & 3.48 & 3.83 & 4.06 & 3.48 & 3.71 & 3.86 & 3.96 & 3.93$\pm$0.10\\
AsF $^3\Sigma^-$      & 3.88 & 4.13 & 4.30 & 3.91 & 4.16 & 4.32 & 3.85 & 4.08 & 4.22 & 4.2 & 4.21\\
ditto aug-cc          &      &      &      & 4.12 & 4.26 & 4.37 \\
AsH $^3\Sigma^-$      & 2.63 & 2.72 & 2.79 & 2.63 & 2.73 & 2.79 & 2.64 & 2.71 & 2.76 & 2.76 (AsD)
& 2.80$\pm$0.03 \\
AsN                   & 4.48 & 4.85 & 5.09 & 4.54 & 4.90 & 5.14 & 4.47 & 4.72 & 4.88 & --- & 5.03$\pm$0.02\\
AsO $^2\Pi$           & 4.58 & 4.90 & 5.11 & 4.63 & 4.94 & 5.14 & 4.56 & 4.80 & 4.97 & $\leq$4.98& 4.95$\pm$0.08\\
AsP                   & 3.90 & 4.24 & 4.46 & 3.94 & 4.27 & 4.49 & 3.90 & 4.16 & 4.33 & --- & 4.45 \\
AsS $^2\Pi$           & 3.47 & 3.79 & 3.99 & 3.51 & 3.81 & 4.02 & 3.50 & 3.73 & 3.88 & --- & 3.90$\pm$0.07\\
BBr                   & 4.23 & 4.32 & 4.38 & 4.25 & 4.34 & 4.40 & 4.18 & 4.28 & 4.34 & $\leq$4.49$^a$ & 4.07\\
ditto aug-cc          &      &      &      & 4.26 & 4.34 & 4.40 & 4.21 & 4.29 & 4.35\\
Br$_2$                & 1.58 & 1.80 & 1.94 & 1.61 & 1.82 & 1.96 & 1.69 & 1.84 & 1.93 & 1.9707 & idem \\
ditto aug-cc          &      &      &      & 1.76 & 1.88 & 1.96 & 1.75 & 1.87 & 1.94 \\
BrCl                  & 1.90 & 2.08 & 2.20 & 1.91 & 2.09 & 2.21 & 1.94 & 2.10 & 2.20 & 2.233 & 2.223$\pm$0.003\\
ditto aug-cc          &      &      &      & 2.01 & 2.13 & 2.21 \\
BrF                   & 2.18 & 2.40 & 2.54 & 2.20 & 2.41 & 2.55 & 2.25 & 2.44 & 2.57 & 2.548 & 2.87$\pm$0.12\\
ditto aug-cc          &      &      &      & 2.41 & 2.50 & 2.56 \\
CSe                   & 5.72 & 5.92 & 6.06 & 5.76 & 5.97 & 6.06 & 5.73 & 5.89 & 6.00 & 5.98 & 6.08$\pm$0.06\\
HBr                   & 3.62 & 3.69 & 3.74 & 3.64 & 3.71 & 3.76 & 3.63 & 3.70 & 3.75 & 3.758 & idem\\
ditto aug-cc          & 3.66 & 3.71 & 3.74 & 3.68 & 3.73 & 3.76 & 3.67 & 3.72 & 3.75 &       \\
NSe $^2\Pi$           & 3.23 & 3.51 & 3.70 & 3.27 & 3.55 & 3.73 & 3.29 & 3.51 & 3.66 & 4.0 & 3.80$\pm$0.11\\
Se$_2$ $^3\Sigma^-_g$ & 2.94 & 3.19 & 3.35 & 2.99 & 3.23 & 3.39 & 3.05 & 3.23 & 3.34 & 3.410 & 3.417$\pm$0.004\\
SeH                   & 3.08 & 3.16 & 3.22 & 3.09 & 3.18 & 3.23 & 3.11 & 3.19 & 3.24 & ---   & 3.221$\pm$0.01\\
SeO $^3\Sigma^-$      & 4.02 & 4.29 & 4.47 & 4.05 & 4.32 & 4.49 & 4.08 & 4.29 & 4.42 & 4.41 & 4.78$\pm$0.22\\
SeS $^3\Sigma^-$      & 3.42 & 3.65 & 3.80 & 3.45 & 3.68 & 3.83 & 3.46 & 3.65 & 3.78 & 3.7 & 3.81$\pm$0.07\\
SiSe                  & 5.13 & 5.35 & 5.49 & 5.16 & 5.38 & 5.53 & 5.14 & 5.34 & 5.47 & 5.64 & 5.54$\pm$0.13\\
ditto aug-cc          & 5.16 & 5.36 & 5.49 & 5.20 & 5.40 & 5.53 & 5.18 & 5.36 & 5.48 &      \\

\end{tabular}

(a) Predissociation: Ref.\cite{Hub79} notes a possible potential hump of up to 0.13 eV in the upper $a~^1\Pi$ 
state.
\end{table}

\begin{table}
\caption{Binding energies ($D_e$ in eV) for molecules containing fourth row atoms. \label{row4De}}
\squeezetable
\begin{tabular}{lc*{8}{d}}
molecule& \multicolumn{3}{c}{SDB} & \multicolumn{3}{c}{CPP} &  \multicolumn{2}{c}{Experiment} \\
        & VTZ & VQZ & $\infty$ & VTZ & VQZ &  $\infty$ &  HH\cite{Hub79} & KS\cite{De_CRC}\\
\hline
AlI    & 3.49 & 3.60 & 3.67 & 3.52 & 3.63 & 3.71 & 3.77 & 3.81$\pm$0.02\\
ditto aug-cc &&      &      & 3.53 & 3.64 & 3.72\\
GaI    & 3.18 & 3.34 & 3.45 & 3.26 & 3.42 & 3.53 & 3.47 & 3.47$\pm$0.10\\
GeTe   & 3.75 & 4.08 & 4.30 & 3.84 & 4.16 & 4.37 & 4.24 & 4.09$\pm$0.03\\
HI     & 2.90 & 2.96 & 3.00 & 2.93 & 2.99 & 3.03 & 3.0541 & idem\\
ditto aug-cc
       & 2.94 & 2.99 & 3.01 & 2.96 & 3.01 & 3.04 &      \\
I$_2$  & 1.07 & 1.25 & 1.37 & 1.13 & 1.30 & 1.42 & 1.54238 & idem\\
ditto aug-cc &&      &      & 1.15 & 1.36 & 1.50 \\
IBr    & 1.37 & 1.57 & 1.70 & 1.42 & 1.61 & 1.74 & 1.817 & idem \\
ditto aug-cc &&      &      & 1.54 & 1.67 & 1.75 \\
ICl    & 1.73 & 1.90 & 2.02 & 1.76 & 1.93 & 2.04 & 2.1531 & idem\\
ditto aug-cc &&      &      & 1.86 & 1.98 & 2.06 \\
IF     & 2.24 & 2.48 & 2.64 & 2.27 & 2.51 & 2.66 & 2.879 & $\leq$2.78\\
ditto aug-cc &&      &      & 2.51 & 2.61 & 2.68 \\
InBr   & 3.76 & 4.03 & 4.21 & 3.85 & 4.13 & 4.32 & 3.99 & 4.27$\pm$0.22\\
 (b)   & 3.73 & 3.85 & 3.93 &      &      &     \\
 (c)   & 3.75 & 3.86 & 3.93 & 3.77 & 3.88 & 3.95\\
InCl   & 4.32 & 4.60 & 4.79 & 4.41 & 4.70 & 4.89 & 4.44 & 4.52$\pm$0.08\\
(c)\cite{Bau99In}& 4.29  & 4.45 & &      &      &\\
(e)\cite{Bau99In}& 4.32  & 4.43 &      &      & 4.46$^f$ &\\
InF    & 5.41 & 5.87 & 6.17 & 5.50 & 5.97 & 6.28 & 5.25 & 5.21$\pm$0.15\\
 (b)   & 5.18 & 5.35 & 5.46 \\
 (c)   & 5.35 & 5.42 & 5.47 \\
 (d)   & 5.26 & 5.42 & 5.53 \\
 (e)   & 5.42 & 5.50 & 5.55 \\
InH    & 2.45 & 2.58 & 2.66 & 2.44 & 2.58 & 2.66 & 2.48 & idem\\
 (b)   & 2.43 & 2.47 & 2.49 \\
 (d)   & 2.46 & 2.50 & 2.53 \\
InI    & 3.06 & 3.29 & 3.44 & 3.17 & 3.40 & 3.55 & 3.43 & 3.41$\pm$0.01\\
 (b)   & 3.07 & 3.16 & 3.22 & 3.11 & 3.20 & 3.26 \\
Sb$_2$ & 2.51 & 2.88 & 3.13 & 2.60 & 2.97 & 3.21 & 2.995 $^a$&3.07$\pm$0.07\\
SbF $^3\Sigma^-$
       & 3.77 & 4.12 & 4.36 & 3.82 & 4.17 & 4.40 & 4.4 &4.5$\pm$0.1 \\
ditto aug-cc       &      &      &      & 4.11 & 4.26 & 4.35\\
SbH $^3\Sigma^-$
       & 2.43 & 2.52 & 2.59 & 2.44 & 2.54 & 2.60 & --- & ---     \\
SbP    & 3.20 & 3.51 & 3.71 & 3.26 & 3.57 & 3.78 & 3.68 & idem\\
SnO    & 4.86 & 5.28 & 5.54 & 4.96 & 5.38 & 5.65 & 5.49 & 5.48$\pm$0.13\\
SnS    & 4.31 & 4.61 & 4.81 & 4.39 & 4.69 & 4.89 & 4.77 & 4.78$\pm$0.03\\
SnSe   & 3.83 & 4.12 & 4.31 & 3.92 & 4.22 & 4.41 & 4.20 &4.13$\pm$0.06\\
ditto aug-cc
       & 3.97 & 4.15 & 4.27 & 4.06 & 4.24 & 4.35 &      \\
SnTe   & 3.29 & 3.56 & 3.74 & 3.39 & 3.66 & 3.84 & 3.69 & idem\\
ditto aug-cc
       & 3.40 & 3.58 & 3.70 & 3.50 & 3.68 & 3.79 &      \\
\end{tabular}
(a) Sb$_2$ ($D_e$): H. Sontag and R. Weber, \jmsp{91}{72}{1982}.\\
(b) using Bauschlicher\cite{Bau99In} cc-pV$n$Z basis sets on In in conjunction 
with cc-pV$n$Z and SDB-cc-pV$n$Z basis sets on other element.\\
(c) ditto, but using `augmented' basis sets on other element.\\
(d) as (b), but using all-electron basis set on In.\\
(e) as (c), but using all-electron basis set on In.\\
(f) This work, correlating (2s2p) electrons on Cl and using the MTavqz core-correlation basis set\cite{w1} on Cl.\\
\end{table}

\begin{table}
\caption{Binding energies ($D_e$ in eV) for polar molecules of Ga and Ge \label{GaGeDe}}
\squeezetable
\begin{tabular}{lc*{8}{d}}
molecule& basis & SDB & CPP &SDB-      & all $e^-$ & all $e^-$ with && \multicolumn{2}{c}{Experiment} \\ 
        &       &     &     &cc-pDV$n$Z& cc-pV$n$Z &cc-pV$n$Z &cc-pDV$n$Z& HH\cite{Hub79} & KS\cite{De_CRC}\\
(3d) corr.&     & no  & no  &yes& no & yes & yes\\
\hline
GaBr    & VTZ      & 3.90 & 3.97 & 3.86 & 3.97 & 4.04 & 4.02 & 4.31 & 4.58$\pm$0.18\\
        & VQZ      & 4.09 & 4.16 & 4.15 & 4.08 & 4.15 & 4.12 & \\
        & $\infty$ & 4.22 & 4.29 & 4.35 & 4.16 & 4.23 & 4.19 & \\
GaCl    & VTZ      & 4.46 & 4.52 & 4.60 & 4.53 & 4.63 & 4.60 & 4.92 & 4.96$\pm$0.13\\
        & VQZ      & 4.66 & 4.72 & 4.73 & 4.65 & 4.75 & 4.72 & \\
        & $\infty$ & 4.79 & 4.84 & 4.81 & 4.73 & 4.83 & 4.80 & \\
GaF     & VTZ      & 5.67 & 5.74 & 5.83 & 5.74 & 5.87 & 5.85 & 5.98 & 5.95$\pm$0.15\\
        & VQZ      & 5.94 & 6.01 & 6.00 & 5.91 & 6.03 & 6.01 & \\
        & $\infty$ & 6.12 & 6.18 & 6.12 & 6.02 & 6.14 & 6.11 & \\
        & AVTZ     &      & 5.95 & 6.01\\
        & AVQZ     &      & 6.08 & 6.09\\
        & $\infty$ &      & 6.17 & 6.15\\
        & DVTZ+2s2p &&& 5.80 &      \\
        & DVQZ+3s3p &&& 5.97 &      \\
        & $\infty$ &&& 6.09 &      \\
GaH     & VTZ      & 2.69 & 2.69 & 2.81 & 2.77 & 2.86 & 2.82 & $<$ 2.84 & $\leq$2.80\\
        & VQZ      & 2.77 & 2.77 & 2.85 & 2.80 & 2.89 & 2.84 & \\
        & $\infty$ & 2.82 & 2.81 & 2.88 & 2.82 & 2.90 & 2.85 & \\
GeF $^2\Pi$
        & VTZ      & 4.78 & 4.83 & 4.99 & 4.83 & 4.99 & 4.98 & 5.0 & 5.0$\pm$0.2\\
        & VQZ      & 5.07 & 5.12 & 5.17 & 5.02 & 5.20 & 5.18 & \\
        & $\infty$ & 5.27 & 5.31 & 5.31 & 5.14 & 5.33 & 5.28 & \\
        & AVTZ     &      & 5.06\\
        & AVQZ     &      & 5.20\\
        & $\infty$ &      & 5.30\\
GeH $^2\Pi$
        & VTZ      & 2.57 & 2.58 & 2.69 & 2.63 & 2.70 & 2.66 & $<$ 3.3 & $\leq$3.3\\
        & VQZ      & 2.65 & 2.66 & 2.72 & 2.68 & 2.76 & 2.72 & \\
        & $\infty$ & 2.71 & 2.71 & 2.75 & 2.72 & 2.80 & 2.76 & \\
GeO     & VTZ      & 6.19 & 6.27 & 6.57 & 6.41 & 6.54 & 6.53 & 6.78 & 6.80$\pm$0.13\\
        & VQZ      & 6.57 & 6.65 & 6.76 & 6.60 & 6.77 & 6.74 & \\
        & $\infty$ & 6.82 & 6.89 & 6.88 & 6.73 & 6.93 & 6.87 & \\
GeS     & VTZ      & 5.10 & 5.16 & 5.36 & 5.25 & 5.35 & 5.31 & 5.67 & 5.50$\pm$0.03\\
        & VQZ      & 5.40 & 5.57 & 5.52 & 5.44 & 5.55 & 5.51 & \\
        & $\infty$ & 5.59 & 5.83 & 5.63 & 5.56 & 5.68 & 5.63 & \\
GeSe    & VTZ      & 4.45 & 4.53 & 4.71 & 4.64 & 4.73 & 4.70 & 4.98$\pm$0.10 & 4.99$\pm$0.02\\
        & VQZ      & 4.79 & 4.86 & 4.89 & 4.81 & 4.91 & 4.87 & \\
        & $\infty$ & 5.01 & 5.07 & 5.01 & 4.92 & 5.03 & 4.98 & \\
\end{tabular}
\end{table}

\begin{table}
\caption{State-averaged optimum $f$ and $g$ exponents for the transition metals, to be used
in conjunction with Stuttgart-Dresden ECPs and [6s5p3d] contracted valence basis sets. The cc-pVTZ
and SDB-cc-pVTZ basis sets are recommended for the other elements\label{transition}}

\begin{tabular}{lcccccccccc}
$(s)^1(d)^{n-1}$ state & $^4F$ & $^5F$ & $^6D$ & $^7S$ & $^6D$ & $^5F$ & $^4F$ & $^3D$ & $^2S$ & N/A\\
$(s)^2(d)^{n-2}$ state & $^2D$ & $^3F$ & $^4F$ & $^5D$ & $^6S$ & $^5D$ & $^4F$ & $^3F$ & $^2D$ & $^1S$\\
 & Sc & Ti & V & Cr & Mn & Fe & Co & Ni & Cu & Zn\\
$\zeta_{f1}$ & 0.180 & 0.285 & 0.425 & 0.640 & 0.795 & 0.871 & 1.019 & 1.182 & 1.315 & 1.498\\
$\zeta_{f2}$ & 0.764 & 1.264 & 1.788 & 2.555 & 3.118 & 3.516 & 4.076 & 4.685 & 5.208 & 5.871\\
$\zeta_g$ & 0.347 & 0.636 & 1.048 & 1.712 & 1.964 & 2.269 & 2.711 & 3.212 & 3.665 & 4.365\\
 & Y & Zr & Nb & Mo & Tc & Ru & Rh & Pd & Ag & Cd\\
$\zeta_{f1}$ & 0.144 & 0.236 & 0.261 & 0.338 & 0.398 & 0.478 & 0.567 & 0.621 & 0.732 & 0.834\\
$\zeta_{f2}$ & 0.546 & 0.883 & 0.970 & 1.223 & 1.430 & 1.666 & 1.989 & 2.203 & 2.537 & 2.853\\
$\zeta_g$ & 0.249 & 0.547 & 0.536 & 0.744 & 0.918 & 1.057 & 1.236 & 1.385 & 1.587 & 1.795\\
 & La & Hf & Ta & W & Re & Os & Ir & Pt & Au & Hg\\
$\zeta_{f1}$ & 0.120 & 0.163 & 0.210 & 0.256 & 0.327 & 0.347 & 0.395 & 0.443 & 0.498 & 0.545\\
$\zeta_{f2}$ & 0.456 & 0.557 & 0.697 & 0.825 & 0.955 & 1.067 & 1.189 & 1.323 & 1.461 & 1.580\\
$\zeta_g$ & 0.209 & 0.352 & 0.472 & 0.627 & 0.636 & 0.860 & 0.982 & 1.100 & 1.218 & 1.384\\
\end{tabular}

Exponents were averaged over $(s)^1(d)^{n-1}$ and $(s)^2(d)^{n-2}$ states, except for Pd where in addition
the $(s)^0(d)^{10}$ ground state was used, and Zn/Cd/Hg for
which only the $(s)^2(d)^{n-2}$ is involved.

\end{table}


\begin{thebibliography}{99}

\bibitem{ccreviews} \oneauth{R. J.}{Bartlett}
\jpc{93}{1697}{1989}; 
\oneauth{P. R.}{Taylor}, 
in {\it Lecture Notes in Quantum Chemistry II} (ed. B. O. Roos),
Lecture Notes in Chemistry {\bf 64}, 125 (1994);
\twoauth{R. J.}{Bartlett}{J. F.}{Stanton} in
{\it Reviews in Computational Chemistry, Vol. V} (Lipkowitz, K. B.,
Boyd, D. B., Eds.) VCH, New York, 1994, pp. 65--169;
\twoauth{T. J.}{Lee}{G. E.}{Scuseria}
\inbook{Quantum mechanical
electronic structure calculations with chemical accuracy}{ed. S.
R. Langhoff}{Kluwer}{Dordrecht, The Netherlands}{1995}, pp. 47--108;
\oneauth{R. J.}{Bartlett} in 
{\it Modern Electronic Structure Theory, Vol. 2} (Yarkony, D. R., Ed.);
World Scientific, Singapore, 1995, pp. 1047--1131.

\bibitem{Hel95} T. Helgaker and P. R. Taylor, in {\it Modern electronic
structure theory, Vol II}, Ed. D.R.Yarkony (World Scientific Publishing,
Singapore, 1995).
\bibitem{Alm87} J. Alml\"{o}f and P. R. Taylor,
\JCP{86}{4070}{1987} 
\bibitem{WMR}\auth{P. O.}{Widmark} \auth{P. \AA.}{Malmqvist} \andauth{B. O.}{Roos}
\tcaold{77}{291}{1990};
\auth{P. O.}{Widmark} \auth{B. J.}{Persson} \andauth{B. O.}{Roos}
\tcaold{79}{419}{1991};
\auth{K.}{Pierloot} \auth{B.}{Dumez}
\auth{P. \AA.}{Malmqvist} \andauth{B. O.}{Roos}
\tcaold{90}{87}{1995};
\auth{R.}{Pou-Amerigo} \auth{M.}{Merchan} \auth{I.}{Nebot-Gil}
\auth{P. O.}{Widmark} \andauth{B. O.}{Roos}
\tcaold{92}{149}{1995}.


\bibitem{Dun89} \auth{T. H.}{Dunning Jr.} \JCP{90}{1007}{1989}.
\bibitem{DunECC}T.H. Dunning, Jr., K.A. Peterson, and D.E. Woon, ``Basis sets: correlation
consistent", in {\it Encyclopedia of Computational Chemistry}, ed. P. von Ragu\'e Schleyer (Wiley \& Sons, 1998), vol. {\bf 1}, pp.  88--115.
\bibitem{Par89}R. G. Parr and W. Yang, \book{Density functional theory for 
atoms and molecules}{Oxford University Press}{Oxford}{1989}
\bibitem{And97} A. C. Scheiner, J. Baker, and J. W. Andzelm , \jcc{18}{775}{1997}
\bibitem{MarVUB} J. M. L. Martin, \inbook{Density Functional Theory --- a bridge between chemistry and physics}{P. Geerlings, F. De Proft, and W. Langenaeker}{VUB Press}{Brussels}{2000}
\bibitem{DePVUB} F. De Proft, \inbook{Density Functional Theory --- a bridge between chemistry and physics}{P. Geerlings, F. De Proft, and W. Langenaeker}{VUB Press}{Brussels}{2000}
\bibitem{ip} F. De Proft and P. Geerlings, \JCP{106}{3270}{1997}
\bibitem{quad} F. De Proft, F. Tielens, and P. Geerlings, \theochem{506}{1}{2000}

\bibitem{Bau95} \twoauth{C. W.}{Bauschlicher Jr.}{P. R.}{Taylor} \tcaold{86}{13}{1993}; \oneauth{C. W.}{Bauschlicher Jr.}
\tcaold{92}{183}{1995}

\bibitem{Wil99} \auth{A. K.}{Wilson} \auth{D. E.}{Woon} \auth{K. A.}{Peterson} \andauth{T. H.}{Dunning Jr.}
\JCP{110}{7667}{1999}

\bibitem{Rh-PCP} A. Sundermann, O. Uzan, D. Milstein, and J. M. L. Martin, \jacs{122}{7095}{2000}
\bibitem{Rh-PCN} A. Sundermann, O. Uzan, and J. M. L. Martin, {\it J. Am. Chem. Soc.}, submitted.
\bibitem{Heck} A. Sundermann, O. Uzan, and J. M. L. Martin, {\it Chemistry --- a European Journal}, submitted.

\bibitem{Bak95} J. Baker, M. Muir, and
J. Andzelm, \JCP{102}{2063}{1995}; J. Andzelm and P. R. Taylor,
\cpl{237}{53}{1995}.

\bibitem{Tru2K} B. J. Lynch, P. L. Fast, M. Harris, and D. G. Truhlar,
\jpca{104}{4811}{2000}.

\bibitem{sn2} S. Parthiban, G. de Oliveira, and J. M. L. Martin, 
{\it J. Phys. Chem. A}, in press (JP0031000).
\bibitem{Pyk88} P. Pyykk\"o, \CR{88}{563}{1988}

\bibitem{Hess2000} \twoauth{M.}{Reiher}{B. A.}{Hess} 
 \inbook{Modern methods and algorithms of
quantum chemistry}{J. Grotendorst}{NIC Series Vol. 1}{Forschungszentrum
J\"ulich}{2000}
\bibitem{KraSteARPC84} M. Krauss and W.J. Stevens, \arpc{35}{357}{1984}
\bibitem{Ermler1988}W.C. Ermler, R.B. Ross, and P A. Christiansen, \jcite{Adv. Quantum Chem.}{19}{139}{1988}
\bibitem{Gropen1988} O. Gropen, \inbook{Methods in
Computational Chemistry, Vol. 2}{S. Wilson}{Plenum}{New York}{1988}.

\bibitem{Dolg2000} M. Dolg, \inbook{Modern methods and algorithms of
quantum chemistry}{J. Grotendorst}{NIC Series Vol. 1}{Forschungszentrum
J\"ulich}{2000}

\bibitem{lanl}\twoauth{P. J.}{Hay}{W. R.}{Wadt} \JCP{82}{270}{1985};
\twoauth{W. R.}{Wadt}{P. J.}{Hay} \JCP{82}{284}{1985};
\twoauth{P. J.}{Hay}{W. R.}{Wadt} \JCP{82}{299}{1985}.
\bibitem{sbk}{W. J. Stevens, M. Krauss, H. Basch, and P. G. Jasien, \jcite{Can.
J. Chem.}{70}{612}{1992} and references therein.}
\bibitem{ermler-christensen}
\auth{M. M.}{Hurley} \auth{L. F.}{Pacios} \auth{P. A.}{Christiansen}
\auth{R. B.}{Ross} \andauth{W. C.}{Ermler} \JCP{84}{6840}{1986}

\bibitem{sddmain} M. Dolg, U. Wedig, H. Stoll, and H. Preuss, \JCP{86}{866}{1987}
and subsequent papers: see ref.\cite{Dolg2000} or \cite{sddonline} for a complete reference list.

\bibitem{g98}
\auth{M. J.}{Frisch} \auth{G. W.}{Trucks} \auth{H. B.}{Schlegel}
\auth{G. E.}{Scuseria} \auth{M. A.}{Robb} \auth{J. R.}{Cheeseman}
\auth{V. G.}{Zakrzewski} \auth{J. A.}{Montgomery} \auth{R. E.}{Stratmann}
\auth{J. C.}{Burant} \auth{S.}{Dapprich} \auth{J. M.}{Millam}
\auth{A. D.}{Daniels} \auth{K. N.}{Kudin} \auth{M. C.}{Strain}
\auth{O.}{Farkas} \auth{J.}{Tomasi} \auth{V.}{Barone} \auth{M.}{Cossi}
\auth{R.}{Cammi} \auth{B.}{Mennucci} \auth{C.}{Pomelli} \auth{C.}{Adamo}
\auth{S.}{Clifford} \auth{J.}{Ochterski} \auth{G. A.}{Petersson}
\auth{P. Y.}{Ayala} \auth{Q.}{Cui} \auth{K.}{Morokuma} \auth{D. K.}{Malick}
\auth{A. D.}{Rabuck} \auth{K.}{Raghavachari} \auth{J. B.}{Foresman}
\auth{J.}{Cioslowski} \auth{J. V.}{Ortiz} \auth{B. B.}{Stefanov} \auth{G.}{Liu}
\auth{A.}{Liashenko} \auth{P.}{Piskorz} \auth{I.}{Komaromi} \auth{R.}{Gomperts}
\auth{R. L.}{Martin} \auth{D. J.}{Fox} \auth{T.}{Keith} \auth{M. A.}{Al-Laham}
\auth{C. Y.}{Peng} \auth{A.}{Nanayakkara} \auth{C.}{Gonzalez}
\auth{M.}{Challacombe} \auth{P. M. W.}{Gill} \auth{B. G.}{Johnson}
\auth{W.}{Chen} \auth{M. W.}{Wong} \auth{J. L.}{Andres}
\auth{M.}{Head-Gordon} \auth{E. S.}{Replogle} \andauth{J. A.}{Pople}
{\it Gaussian 98, Revision A.7} (Gaussian, Inc., Pittsburgh, PA, 1999).

\bibitem{m2k}
MOLPRO is a package of ab initio programs written by H.-J. Werner and P. J. Knowles, with contributions from
R. D. Amos, A. Bernhardsson, A. Berning,
P. Celani, D. L. Cooper, M. J. O. Deegan, 
A. J. Dobbyn, F. Eckert, C. Hampel, G. Hetzer, 
T. Korona, R. Lindh, A. W. Lloyd, S. J. McNicholas, F. R. Manby,
W. Meyer, M. E. Mura, A. Nicklass, P. Palmieri, R. Pitzer, 
G. Rauhut, M. Sch\"utz, H. Stoll, A. J. Stone, R. Tarroni, and T. Thorsteinsson. 

\bibitem{cpp} 
W. M\"uller, J. Flesch, and W. Meyer, \JCP{80}{3297}{1984};
P. Fuentealba, H. Preuss, H. Stoll, and L. von Szentp\'aly,
\cpl{89}{418}{1982}
P. Schwerdtfeger and H. Silberbach, \PRA{37}{2834}{1988};
\erratum{42}{665}{1990}

\bibitem{Bau99In} C. W. Bauschlicher Jr., \cpl{305}{446}{1999}

\bibitem{Bau99InBIS} \oneauth{C. W.}{Bauschlicher Jr.} \jpca{103}{6429}{1999}

\bibitem{domin} P. Spellucci, {\it domin},  a subroutine for BFGS minimization
(Department of Mathematics, Technical University of Darmstadt, Germany, 1996).

\bibitem{Ber93} A. Bergner, M. Dolg, W. Kuechle, H. Stoll, H. Preuss,
\mph{80}{1431}{1993}.

\bibitem{Woo80} J.H. Wood and A.M. Boring, \PRB{18}{2701}{1978}.

\bibitem{avnz} \auth{R. A.}{Kendall} \auth{T. H.}{Dunning}
\andauth{R. J.}{Harrison} \JCP{96}{6796}{1992}

\bibitem{Hub79} K. P. Huber and G. Herzberg, 
\book{Constants of Diatomic Molecules}{Van Nostrand Reinhold}{New York}{1979}

\bibitem{Dun32} J. L. Dunham, \jcite{Phys. Rev.}{41}{721}{1932}

\bibitem{Wat93} 
\auth{J. D.}{Watts} \auth{J.}{Gauss} \andauth{R. J.}{Bartlett}
\JCP{98}{8718}{1993}

\bibitem{Ige88} G. Igel-Mann, H. Stoll, and H. Preuss, 
\mph{65}{1321}{1988}

\bibitem{Lei97} T. Leininger, A. Berning, A. Nicklass, H. Stoll, H.-J. Werner, 
and H.-J. Flad,
\cp{217}{19}{1997}

\bibitem{sddonline} \url{http://www.theochem.uni-stuttgart.de/pseudopotentials/accueil.html}

\bibitem{Woo93} \twoauth{D. E.}{Woon}{T. H.}{Dunning Jr.} \JCP{98}{1358}{1993}

\bibitem{ch} J. M. L. Martin, \cpl{292}{411}{1998}

\bibitem{Bau98Ga} \oneauth{C. W.}{Bauschlicher Jr.} \jpca{102}{10424}{1998}

\bibitem{Moc98}  \twoauth{Y.}{Mochizuki}{K.}{Tanaka} \tca{99}{88}{1998}

\bibitem{cc} J. M. L. Martin \cpl{242}{343}{1995}

\bibitem{sword} P. Schwerdtfeger, Th. Fischer, M. Dolg, G. Igel-Mann,
A. Nicklass, H. Stoll, and A. Haaland, \JCP{102}{2050}{1995};
T. Leininger, A. Nicklass, H. Stoll, M. Dolg, and P. Schwerdtfeger,
\JCP{105}{1052}{1996}



\bibitem{De_CRC} J. A. Kerr and D. W. Stocker, ``Bond strengths in diatomic molecules'' in
{\it Handbook of Chemistry and Physics, 80th Edition}, CRC Press, Boca Raton, FL, 1999.

\bibitem{atomsplit} The required fine structure data for the first and
second row, as well as for I, were taken from
\auth{M. W.}{Chase Jr.} \auth{C. A.}{Davies}
\auth{J. R.}{Downey Jr.} \auth{D. J.}{Frurip} \auth{R. A.}{McDonald}
\andauth{A. N.}{Syverud} {\it JANAF thermochemical
tables, 3rd edition}, \jpcrd{14}{supplement 1}{1985}.
Fine structure data for Ga, Ge, Se, and Br were 
obtained from \url{http://physics.nist.gov/cgi-bin/AtData/main_asd}
and for In, Sn, and Te from 
\url{http://cfa-www.harvard.edu/amdata/ampdata/kurucz23/sekur.html}.

\bibitem{w1} J. M. L. Martin and G. de Oliveira, \JCP{111}{1843}{1999}

\bibitem{Hal98}\auth{A.}{Halkier} \auth{T.}{Helgaker}
\auth{P.}{J{\o}rgensen} \auth{W.}{Klopper}
\auth{H.}{Koch} \auth{J.}{Olsen} \andauth{A. K.}{Wilson} \cpl{286}{243}{1998}

\bibitem{Dol96} M. Dolg, \mph{88}{1645}{1996}

\bibitem{l4} \auth{J. M. L.}{Martin} \cpl{259}{669}{1996}

\bibitem{Vis1} \auth{L.}{Visscher} \auth{J.}{Styszynski} \andauth{W. C.}{Nieuwpoort} \JCP{105}{1987}{1996}

\bibitem{Vis2} \twoauth{L.}{Visscher}{K. G.}{Dyall} \JCP{104}{9040}{1996}

\bibitem{Vis3} \auth{W. A.}{de Jong} \auth{J.}{Styszynski} \auth{L.}{Visscher} \andauth{W. C.}{Nieuwpoort}
\JCP{108}{5177}{1998}

\bibitem{relccsd} \auth{L.}{Visscher} \auth{T. J.}{Lee} \andauth{K. G.}{Dyall}
\JCP{105}{8769}{1996}

\bibitem{Pis95} L. Pisani and E. Clementi, \JCP{103}{9321}{1995}

\bibitem{Dol87} M. Dolg, U. Wedig, H. Stoll, and H. Preuss, \JCP{86}{866}{1987}
\bibitem{And90} D. Andrae, U. Haeussermann, M. Dolg, H. Stoll, H. Preuss, \tcaold{77}{123}{1990}
\end{thebibliography}
\end{document}